\newif\ifHeader
\Headertrue 

\documentclass[a4paper]{WileySev}

\usepackage{w-bookps}
\usepackage{mathptmx}
\usepackage[utf8]{inputenc}
\usepackage{latexsym}
\usepackage{amsmath}
\usepackage{amsfonts}
\usepackage{graphicx}

\usepackage{lipsum}
\usepackage{caption}
\usepackage{subcaption}
\usepackage{float}
\usepackage{colortbl}
\usepackage{xcolor}
\usepackage{multicol}
\usepackage{rotating}
\usepackage[sectionbib,square]{natbib}
\usepackage{url}

\usepackage{multind} 

\makeindex{topic}
\makeindex{author}

\docropmarks 

\newcommand{\clock}{\count254=\time \divide\count254 by 60
 \count255=\count254 \multiply\count255 by -60
 \advance\count255 by \time
 \ifnum\count254<10 0\fi\number\count254\,:\,%
 \ifnum\count255<10 0\fi\number\count255}

\newcommand{\rot}{\rotatebox}

\newcommand{\href}[2]{\texttt{#2}}

\setcitestyle{numbers} 
\makeatletter 
\renewcommand\@biblabel[1]{#1.} 
\makeatother
\begin{document}
\bibliographystyle{abbrvnat}

\booktitle{Advances in\\ Network Clustering\\ and Blockmodeling}
\subtitle{}

\authors{%
Patrick Doreian\\ \affil{University of Pittsburgh}
Vladimir Batagelj\\ \affil{University of Ljubljana}
Anu{\v{s}}ka Ferligoj\\ \affil{University of Ljubljana}
}

\ifHeader \titlepage \fi

\offprintinfo{Advances in Network Clustering and Blockmodeling, \\
\today \quad \clock 
}{P. Doreian, V. Batagelj, A. Ferligoj}



\ifHeader
\fi

\setcounter{chapter}{13} \setcounter{page}{250}


\chapter[Scientific co-authorship networks]
{Scientific co-authorship networks}\label{ch:scicol}

\chapterauthors{Marjan Cugmas, Anuska Ferligoj, and Luka Kronegger
\chapteraffil{FDV, University of Ljubljana}}

\section{Introduction}

Network studies of science offer researchers a great insight into the dynamics of knowledge creation and the social structure of scientific society. The flow of ideas and overall cognitive structure of the scientific community is observed through citations between scientific contributions, usually manifested as patents or papers published in scientific journals. The social structure of this society consists of relationships among scientists. {De Haan \cite{de1997authorship} suggests six operationalized indicators of collaborative relations between scientists: co-authorship; shared editorship of publications; shared supervision of PhD projects; writing a research proposal together; participation in formal research programs; and shared organization of scientific conferences.

Due to accessibility and the ease of acquiring data through bibliographic databases, most scientific collaboration analyses are performed on co-authorship data, which play a particularly important role in research into the collaborative social structure of science. Co-authorship networks are personal networks in which the vertices represent authors, and two authors are connected by a tie if they co-authored one or more publications. These ties are necessarily symmetric. The study of community structures through scientific co-authorship is especially important because scientific (sub)disciplines can often display local properties that differ greatly from the properties of the scientific network as a whole.
Co-authorship data have some flaws. The wide pallet of relationships among scientists do not result in common publications \citep{katz1997a, melin1996, laudel2002}. Laudel \cite{laudel2002} reports that about half of scientific collaborations are invisible in formal communication channels because they do not lead to either co-authored publications or formal acknowledgments in scientific texts. On the other hand, we also know that co-authorship sometimes represents false positive relations arising from resource-related issues \citep{ponomariov2016}. Despite this, co-authorship data compromise between quality and cost for an indicator of scientific collaboration.

The study of co-authorship networks has been influenced by the development of quantitative methodological approaches \citep{mali2010}. The choice of relatively simple descriptive statistics, deterministic modeling, stochastic agent-based modeling of network dynamics, or any other method is based on a particular study's objective. In general, which of the many approaches to studying co-authorship networks is chosen depends on the objective of the study under consideration. The most fundamental approaches to studying co-authorship networks relate to co-authorship networks' basic descriptive statistics, such as measuring the number and size of components in the network along with the degree and different measures of closeness and centrality. Some researchers used  an Exponential Random Graph Modeling to test a small world structure \cite{kronegger2012collaboration} in co-authorship networks while some other also proposed studying transformed co-authorship networks where the nodes are articles (instead of researchers) and links between two articles exist if they have one or several of the same authors \cite{gasko2016new}. In the current chapter, we focus on blockmodeling co-authorship networks as a deterministic approach to network analysis. 

There are relatively few applications of blockmodeling to co-authorship networks. This may be due to the method's limitations regarding the size of analyzed networks. One of the earliest applications can be found in \cite{Ferligoj2009}, who compared the results of blockmodeling (clustering of relational data) of a co-authorship network of Slovenian sociologists and the results of clustering with a relational constraint (clustering of attribute and relational data) on the same network according to researchers' publication performance. As expected, the methods produced different results, indicating their use should depend on the research problem under study. The unexpected result of their presented analysis was a core-periphery structure, with seven cores and a periphery, obtained when blockmodeling the co-authorship network. 

Further investigation of the multicore-periphery structure was presented in \cite{kronegger2011} where the authors analyzed the development of a network structure over time. In their analysis of the co-authorship networks of four scientific disciplines (physics, mathematics, biotechnology and sociology) measured in four consecutive 5-year time spans, they observed a multicore-periphery structure was present from early on in the development of each scientific discipline. They also found that, although the number of cores increases with the growth of a discipline, the cores' sizes did not change. The structure's description as constituting multiple cores and a periphery was extended with two elements: a weakly connected semi-periphery, a completely empty periphery and bridging cores, describing clusters of authors connecting two or more cores from the central part of the network. The authors described four levels of network complexity in the network structure's evolution through time:   

\begin{enumerate}
  \item Simple core-periphery form: Simple cores, semi-periphery, periphery
  \item Weakly consolidated core-periphery form: Simple cores, bridging individuals, semi-periphery, periphery
  \item Consolidated core-periphery form: Simple cores, bridging cores, semi-periphery, periphery
  \item Strongly consolidated core-periphery form: Simple cores, bridging cores, bridging individuals, semi-periphery, periphery
\end{enumerate}

The multi-core--semi-periphery--periphery structure was also confirmed in a relatively small co-authorship network constructed from the curricula vitae (CVs) and bibliographies of teaching staff at the Faculty of Humanities and Education Science's Department of Library Science (DHUBI) at the National University of La Plata, Argentina \citep{chinchilla2012}.

Besides describing the overall structure, \citep{kronegger2011} attempted the first (visual) attempts to follow individual units in blockmodels' transition between timespans in order to pinpoint differences in the network dynamics between analyzed disciplines. 
\section{Methods}

A lot of attention has been paid to studying the relationship between collaboration on one side and the quality of research and speed of diffusion of scientific knowledge on the other \citep{hollis2001, frenken2005, abbasi2011, lee2005}. While much research has considered the structure of co-authorship blockmodels \citep{ferligoj2015, moody2004, abbasi2012}, not so much has examined the stability of long-term collaborations. 

Here, it will be illustrated how blockmodeling can be used to reveal the global structure of co-authorship networks and how the stability of the blockmodels obtained can be operationalized and measured. This is especially important when seeking to explain the stability of research teams using common statistical methods such as linear regression.

\subsection{Blockmodeling} \label{blockmodeling}

The goal of blockmodeling is to reduce a large, complex network to a smaller, comprehensible, and interpretable structure \cite{doreian2005generalized}. It can not only be used to find groups of highly linked units in a network, but also the relationships between the groups. While it can reveal a lot of information about the global co-authorship structure, obtaining the solution (especially in the case of direct blockmodeling) can be very computationally expensive where networks with a higher number of units are involved.

The blockmodeling can be either direct or indirect. Indirect blockmodeling is based on a dissimilarity matrix among units. The calculated dissimilarity measure has to be consistent with a chosen equivalence between units. In the studies by Kronegger et al. \cite{kronegger2011} and Cugmas et al. \cite{cugmas2015}, the corrected Euclidean distance, which is consistent with structural equivalence \citep{batagelj1992b}, was used. The process of hierarchical clustering of units can be visualized in a dendrogram in which the units (or groups) and the dissimilarity between the units (or groups) are represented. Kronegger et al. \cite{kronegger2011} and Cugmas et al. \cite{cugmas2015} defined the number of positions based on such visualization.

On the other hand, unlike indirect blockmodeling direct blockmodeling can be achieved through a local optimization procedure \cite{batagelj1992direct}, e.g. using an iterative method where for each displacement of a unit from one group into another, the value of the criterion function is calculated, defined as the difference between the ideal and empirical clustering where the ideal clustering has to express a blockmodel's assumed structure. It turns out that this procedure can be very time-consuming if a higher number of units in the network is analyzed. Cugmas et al. \cite{cugmas2015} also report that the algorithm implemented in Pajek has some difficulties detecting very small, structurally equivalent cores, particularly in the case of scientific disciplines with a very large number of researchers. To mitigate these characteristics, they removed the periphery and the structurally equivalent cliques from the network before applying the procedure. They later merged them to obtain the final result.

\subsection{Measuring the stability of the obtained blockmodels} \label{indices}

The main result of blockmodeling is a partition which assigns a researcher to a certain core, semi-periphery, or periphery. In the case of temporal co-authorship networks (where time is seen as a discrete variable), blockmodeling can be applied for each time period separately such that one partition for each time period is obtained\footnote{Along with the methods for generalised blockmodeling of multilevel networks \citep{vziberna2014blockmodeling}, which can also be used for blockmodeling of temporal networks, different versions of stochastic blockmodeling exist for temporal networks \citep{matias2015statistical, xu2013dynamic, xing2010state, airoldi2007combining}.}. A very important characteristic of temporal co-authorship networks is that some researchers (called newcomers) join the network at a later time period and others (called departures) leave the network at the later time period. Besides the presence of newcomers and departures, also the splitting of cores and merging of cores can be seen as separate features that indicate the lower stability of the obtained blockmodels or cores. 

Nevertheless, a combination of different features usually appears simultaneously; a visualization of each feature is presented in Figure~\ref{factors}. Each visualization is divided into two parts: the white rectangles at the top visualize the clusters (which are cores obtained by blockmodeling in the case of co-authorship blockmodels) from the partition $U=\{u_1,...,u_r\}$ obtained on the set of units from the first time period while the black rectangles on the bottom visualize the clusters from the partition $V=\{v_1,...,v_c\}$ obtained on the set of units from the second time period. Gray rectangles are added to the clusters and visualize the out-comers and newcomers. The links between the rectangles visualize the clusters' stability.

\begin{figure}[!]
    \centering
      \caption{The features that can be used as indicators of less stable clusterings}
      \label{factors}
      \includegraphics[width=1\textwidth]{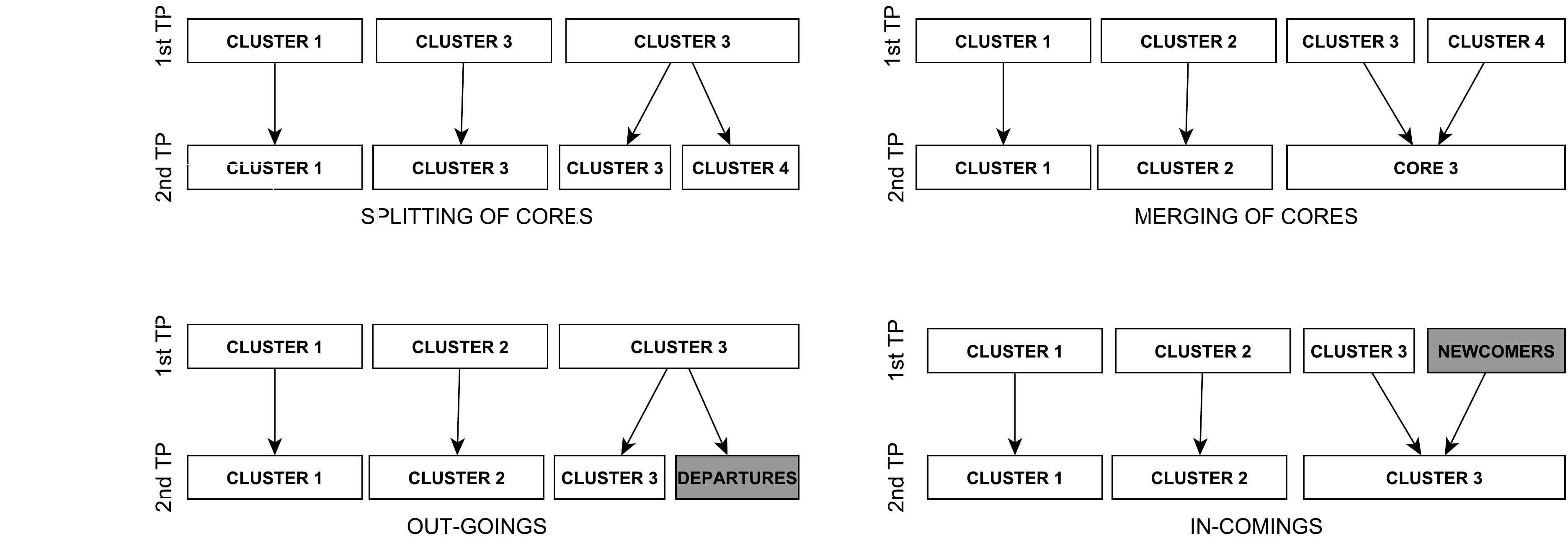}
\end{figure}

\textbf{Adjusted Rand Index} On the two assumptions that the merging and splitting of clusters are indicators of a lower level of cluster stability in time and that there are no newcomers or no departures present (or, at least, that they are neglected), one of the most widely and popular indexes for comparing partitions, the Adjusted Rand Index \cite{hubertarabie1985, saporta2002}, can be used. Here, the adjective "adjusted" refers to the necessary correction for chance since the expected value is usually not 0 in the case of two random and independent partitions. This correction allows the values of the index obtained from different partitions to be compared. Let us focus on the Rand Index \citep{rand1971}, which is defined as 

$$RI=\frac{a+d}{a+b+c+d}$$

\noindent where $a$ stands for the number of pairs of researchers classified in the same cluster in both time periods, $b$ stands for the number of pairs of researchers classified in the same clusters in the first period but in different clusters in the second period, $c$ stands for the number of researchers classified in different clusters in the first, but in the same cluster in the second period and, finally, $d$ stands for the number of pairs of researchers classified in different clusters in both the first and second time periods. Following this definition, the Rand Index can be interpreted in the co-authorship network context as the proportion of all possible pairs of researchers classified in the same or in different clusters in both time periods out of all possible pairs of researchers.

\textbf{Wallace indices} There are situations when the merging and splitting of clusters has to be considered differently. Therefore, one of two Wallace Indices can be used: in the case of the Wallace Index' (WI'), only the splitting of clusters is considered a feature indicating lower cluster stability while with the Wallace Index'' (WI'') only the merging of cores is considered a feature indicating the lower stability of clusters. Formally, WI' is defined as 

$$WI'=\frac{a}{a+b}$$

\noindent where $a$ and $b$ are defined the same as in the case of RI. WI' can be interpreted as the proportion of all researcher pairs placed in the same core in the first period out of the number of all possible researcher pairs placed in the same core in both time periods. Similarly, WI'' is defined as

$$WI''=\frac{a}{a+c}$$

\noindent and interpreted as the ratio between the number of all possible researcher pairs classified in the same cluster in both periods and the number of all possible researcher pairs classified in the same cluster in the second period (the probability that a pair of researchers will be placed in the same cluster in the second period if they were placed in the same cluster in the first period).

\textbf{Modified Rand Index and Wallace indices} As mentioned, it is common in temporal co-authorship networks that some researchers join the network and some leave the network in later time periods. When this happens, one can either simply ignore those researchers when calculating the Rand or Wallace indices, or treat the newcomers and departures as features indicating a lower level of stability of the cores. When the latter is assumed, one has to form new partitions $U'=\{u_1, u_2, ... , u_{r+1}\}$ and $V'=\{v_1, v_2, ... , v_{c+1}\}$ with the new clusters of newcomers $u_{r+1}$ and departures $v_{c+1}$ added to the partitions $U$ and $V$. Then, the Modified Adjusted Rand Index (MARI), the Modified WI', and the Modified WI'' are calculated in the same way as RI, WI', and WI'' where the values in the numerator consider the partitions $U'$ and $V'$. The modified Rand Index and the modified Wallace indices can be further modified in such a way that only newcomers or only departures are considered as features indicating lower core stability (for more details, see \cite{cugmasferligoj2017}) (see Figure~\ref{indices}). 

Along with the modified Rand Index and the modified Wallace indices, Cugmas and Ferligoj \cite{cugmasferligoj2017} proposed a correction for chance (based on Monte Carlo simulations) that allows one to compare the values of indices obtained in different scientific disciplines. With non-adjusted indices, the number of clusters (cores, newcomers, and departures) and the number of researchers also influence the expected value of an index in the case of two random and independent partitions. The expected value of two random and independent partitions in the case of adjusted indices equals 0, and the maximum value of an index is 1. It should be highlighted that higher values of the presented indices indicate a higher level of cluster stability, while lower values indicate a lower level of stability. 

\begin{figure}[!]
    \centering
      \caption{The indices for measuring the stability of cores in time (in brackets the features that lower the stability are given)}
      \label{indices}
      \includegraphics[width=\textwidth]{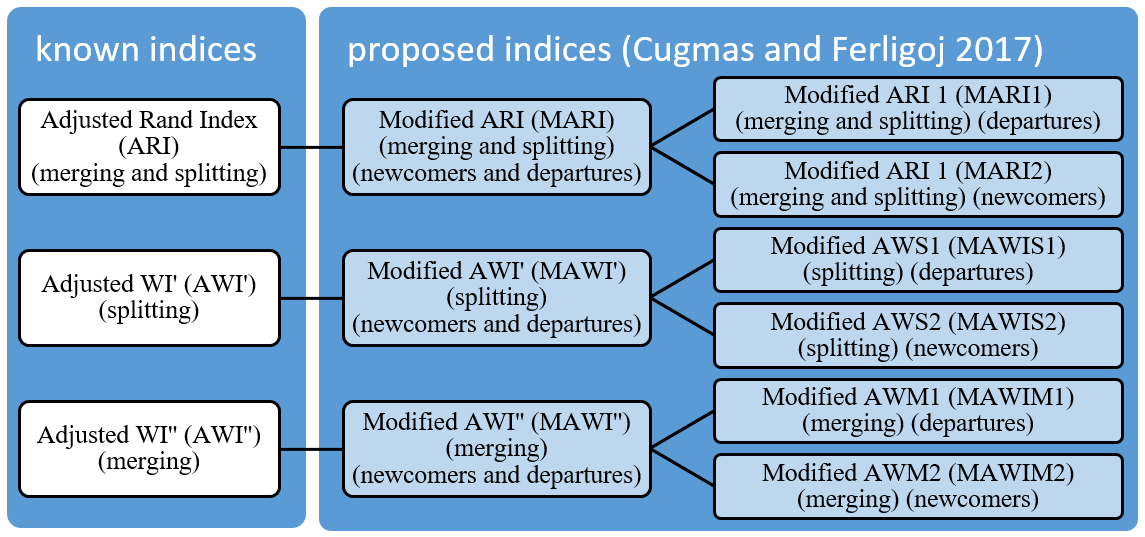}
\end{figure}

\section{The data}

The data for this research were obtained from the Co-operative Online Bibliographic Sy-stem and Services (COBISS) and the Slovenian Current Research Information System (SICRIS) maintained by the Institute of Information Science (IZUM) and the Slovenian Research Agency (SRA). 

SICRIS provides data about all researchers which have an ID assigned by the SRA, including their educational background and field of research according to the SRA's classification scheme. There are 7 scientific fields and 72 scientific disciplines defined in this classification scheme. The 7th scientific field is Interdisciplinary Studies and is not included in the analysis since it has never gained full recognition as a separate research field in Slovenia \cite{ferligoj2015}.

The analyzed data are based on complete personal bibliographies of each researcher (constructed based on SICRIS and COBISS). The network boundaries are therefore defined only by those researchers registered as a researcher at the SRA. Among such researchers, those who published at least one scientific bibliographic unit between 1990 and 2010 are analyzed. The bibliographic units considered as a scientific publication by the SRA are listed in Table~\ref{sci_units}.

\begin{table}[!]
\caption{The number of published scientific bibliographic units by type for two time periods}
\label{sci_units}
\centering
\small
\begin{tabular}{lrr}
\hline
Type of scientific bibliographic unit                                           & 1991 - 2000           & 2001 - 2010           \\ 
                                                                                & (independance)          &  (joining to the EU)                          \\ \hline
Original scientific article                                                       & 26531                 & 47905                 \\
Review article                                                     & 4895                  & 5738                  \\
Short scientific article                                                    & 969                   & 2530                  \\
Published scientific conference contribution (invited lecture)                                     & 3427                  & 5279                  \\
Published scientific conference contribution                                   & 28670                 & 41138                 \\
Independent scientific component part in monograph                                         & 6417                  & 14759                 \\
Scientific monograph                                                        & 1725                  & 2912                  \\
Scientific or documentary films, sound or video recording                               & 44                    & 133                   \\
Complete scientific database or corpus                                       & 73                    & 182                   \\
Patent                                                                          & 381                   & 710                   \\ \hline
Total                                                                           & 73132                 & 121286                \\ \hline 
\end{tabular}
\end{table}

There were 73,132 scientific bibliographic units published between 1991 and 2000 and 121,286 scientific units between 2001 and 2010. The most common are published scientific conference contributions and original scientific articles. The distribution of different types of bibliographic units varies among scientific disciplines. For example, published scientific conference contributions are very common to scientific disciplines from the technical sciences while original scientific articles are frequent among scientific disciplines within the social sciences and humanities. There are differences at the level of scientific disciplines according to the distribution of types of scientific bibliographic units which can be published by one or several researchers. Kronegger et al. \cite{kronegger2015b} who studied the differences between scientific disciplines according to collaboration patterns in time confirmed the scientific discipline geography is more similar to scientific disciplines in the scientific fields natural sciences and mathematics than the scientific field of the humanities where it belongs according to the SRA's classification scheme. Even within a number of scientific disciplines one can expect some differences in types of co-authorships. In the case of sociology, \cite{moody2004} concluded that quantitative work is more likely to be co-authored than non-quantitative work.

Compared to the analysis conducted by Kronegger et al. \cite{kronegger2011} who studied four selected scientific disciplines in four time periods, the current analysis is performed on data for two consecutive 10-year periods between 1991 and 2010. The difference in the length of the periods mainly affects the size and density of the generated co-authorship networks and, in terms of the stability of research teams, result in a lower level of stability. However, the two periods selected reflect a time of major changes to scientific research and development policies in Slovenia. The first period is marked by the independence of Slovenia, meaning that Slovenia had started adopting and implementing its own science policies, while the second period is marked by the country joining the European Union and adopting European Union standards. By the end of this period, Slovenia had already partly integrated its national science system into the European one. 

Although many co-authorship networks are analyzed in this study, we present sociology co-authorship networks as an example. The units represent researchers and a link between two researchers exists if they published at least one scientific bibliographic unit in co-authorship. Therefore, only symmetric links are possible in the case of co-authorship networks. There are also some researchers without any link which are later classified in the so-called periphery, explained in detail in the next section. However, it should be pointed out that the absence of links is not necessarily the consequence of only single-authored scientific bibliographic units by a certain researcher, but can also be the outcome of co-authoring only with researchers who do not have a researcher ID, for example with researchers from abroad. Isolated researchers are present in both time periods. The next important network characteristic which is common to almost all scientific disciplines is that the co-authorship networks grow in time. 

\section{The structure of obtained blockmodels}

Based on four scientific disciplines, Kronegger et al. \cite{kronegger2011} showed that the structure of co-authorship networks consists of the multi-core, semi-periphery, and periphery. To confirm that this structure is also present in other scientific disciplines, Cugmas et al. \cite{cugmas2015} used indirect blockmodeling to detect the approximate number of cores and direct blockmodeling to obtain the final solution as described in Section \ref{blockmodeling}. The assumed blockmodel structure was confirmed in all scientific disciplines included in the analysis. Most disciplines that were excluded (in Figure~\ref{all_disc} indicated by asterisks) were removed due to a small number of researchers in the first or second time period or absence of co-authorship in the current period. One such discipline is theology that did not have a single co-authored scientific bibliographic item published in the first period. It can also be seen in Figure~\ref{all_disc} that the number of researchers who published at least one scientific bibliographic item is increasing over time in almost all scientific disciplines. The average growth in the number of researchers publishing at least one scientific bibliographic item in the second period is 34~\%. Only in the disciplines veterinary medicine, stomatology and mining and geotechnology is a decrease in the number of researchers from the first to the second period observed.

\begin{figure}
    \centering
      \caption{List of scientific disciplines with number of researchers in the first and second periods (an asterisk indicates scientific disciplines which were not considered in the analysis due to a small number of researchers in the first or second time period or absence of co-authorship in the current period)}
      \label{all_disc}
      \includegraphics[width=0.85\textwidth]{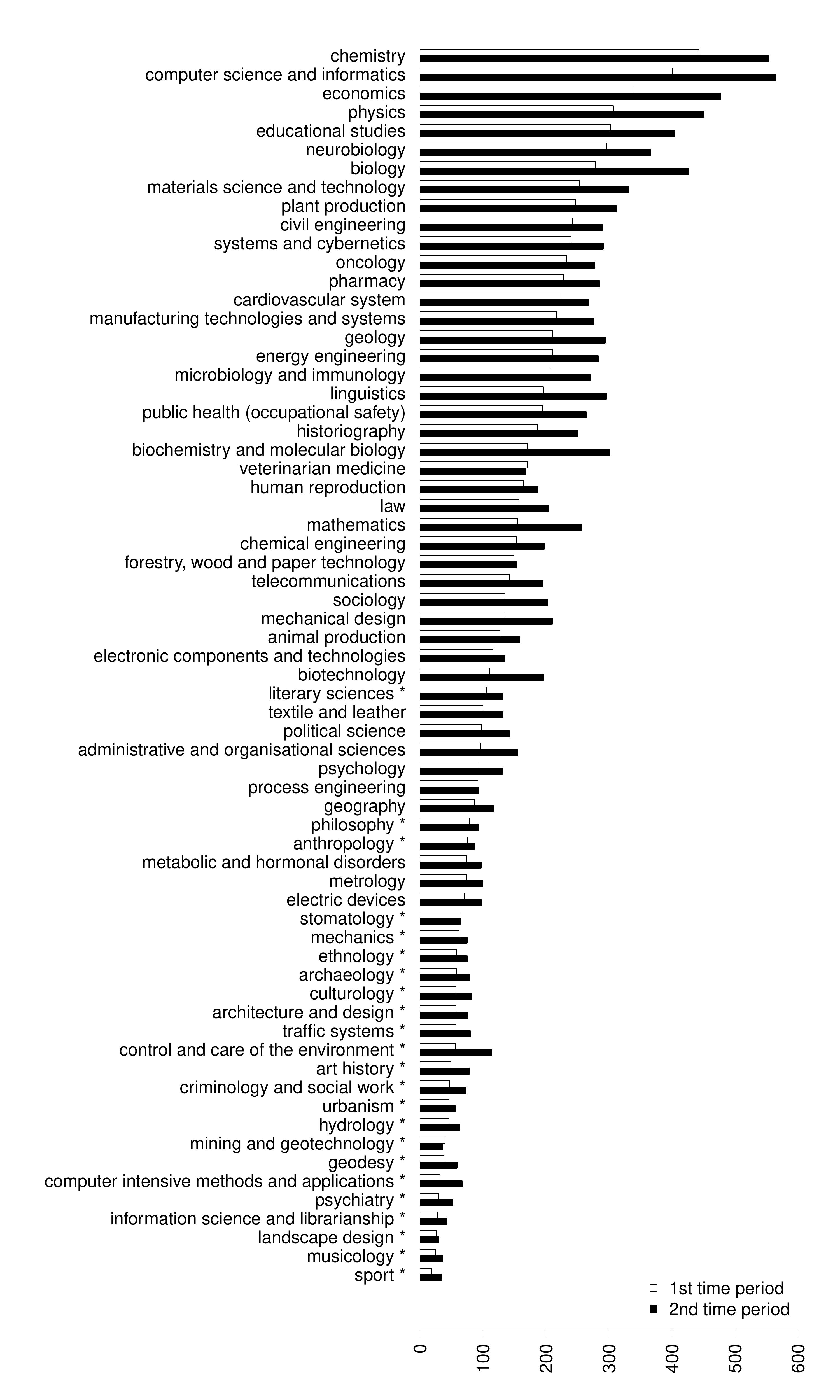}
\end{figure}

Figure~\ref{soc12} visualizes two empirical blockmodels of the scientific discipline sociology. The first blockmodel corresponds to the first period while the second blockmodel corresponds to the second period. The rows and columns of each blockmodel contain the IDs of the researchers, where the black dots in the cells denote co-authorships between two given researchers. A clear multi-core-semi-periphery-periphery structure can be seen in the case of sociology in both time periods. Along with the already described multi-core, semi-periphery, and periphery, in the blockmodel in the first period a so-called bridging core is seen (as a full off-diagonal block) (Figure~\ref{soc1}). The bridging core is a group of researchers who collaborate between each other very systematically and also with researchers from at least two other cores. They are called ``bridging'' since they connect two or more cores. They are relatively common in Slovenian scientific disciplines. There was a minimum of one bridging core in at least one time period in 20 of all analyzed scientific disciplines. 

\begin{figure}[h!]
    \caption{Structure of sociology co-authorship blockmodel for the first and second time periods}
    \label{soc12}
    \centering
    \begin{subfigure}[b]{0.48\textwidth}
        \includegraphics[width=\textwidth]{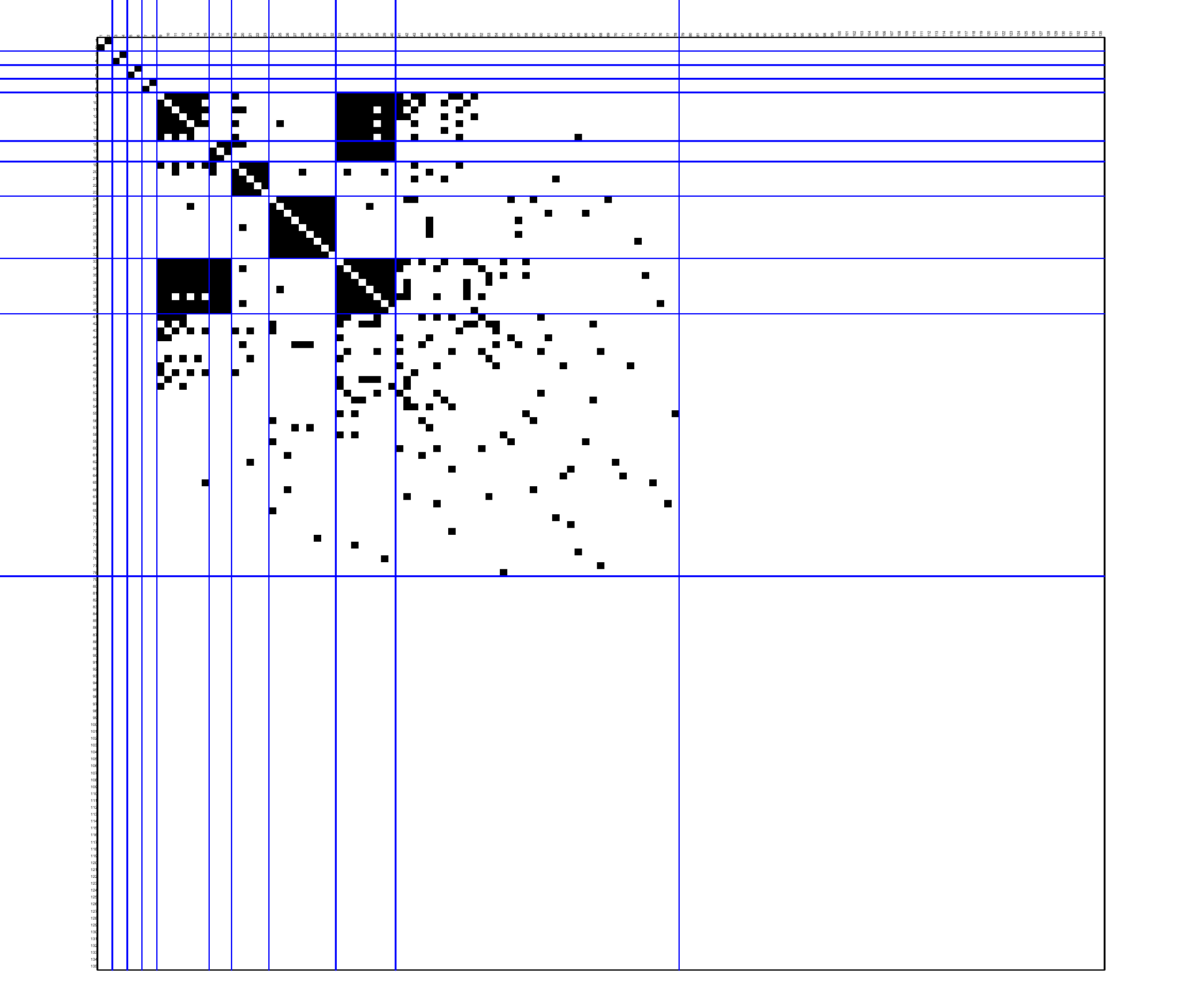}
        \caption{1991 to 2000}
        \label{soc1}
    \end{subfigure}
    \hfill
    \begin{subfigure}[b]{0.48\textwidth}
        \includegraphics[width=\textwidth]{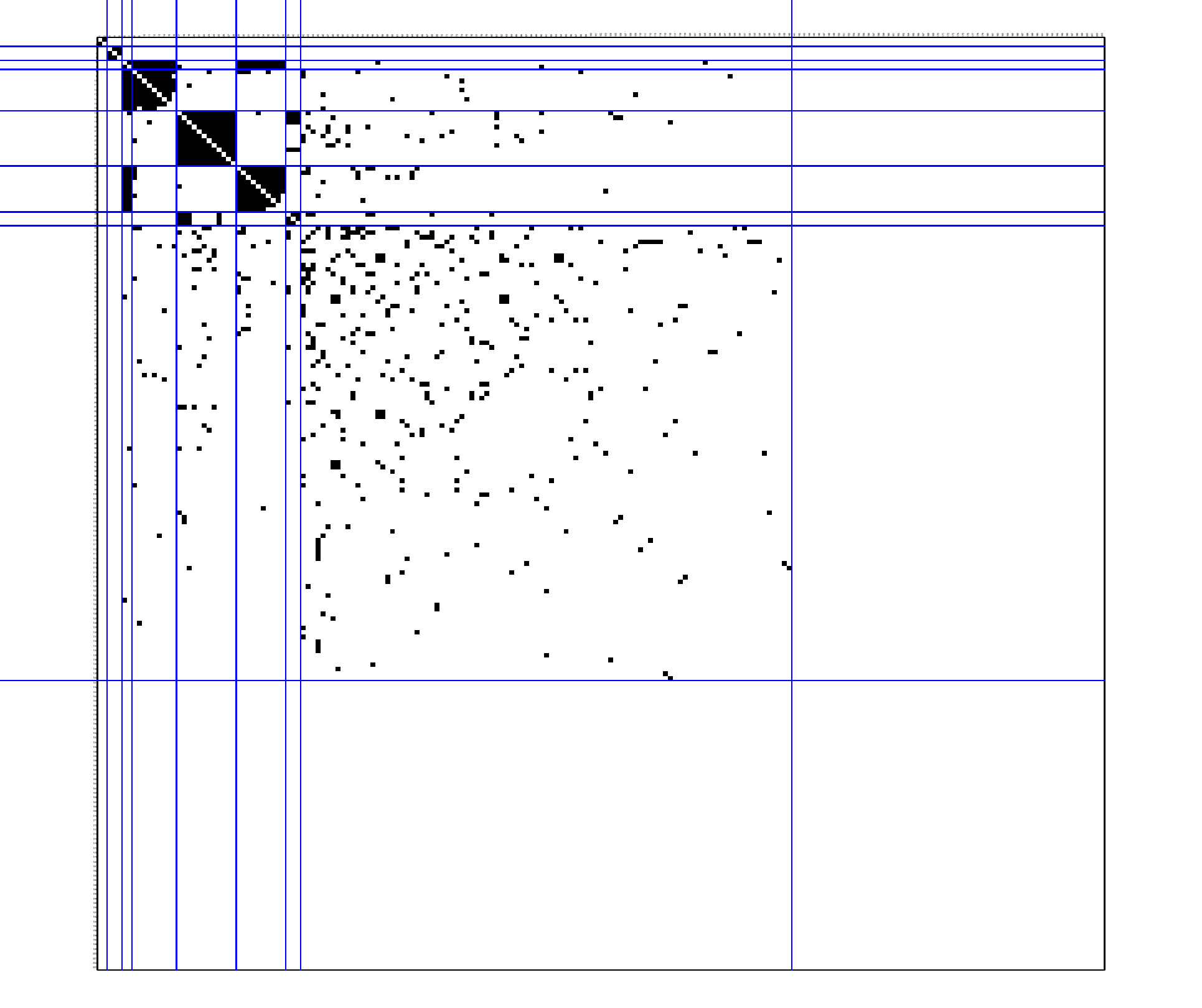}
        \caption{2001 to 2010}
        \label{soc2}
    \end{subfigure}
\end{figure}

The visualization in Figure~\ref{soc12_preh} emphasizes the transitions of researchers between the obtained cores (including the semi-periphery and periphery) for the two periods: the upper part visualizes the classification of researchers for the first period and the bottom part visualizes the classification of researchers for the second period.  It is shown on Figure~\ref{soc12} that the share of researchers classified in the periphery is decreasing in sociology, which cannot be seen in the visualization of researchers' transitions in time in Figure~\ref{preh_a}. This is caused by the newcomers and departures. Figure~\ref{preh_a} reveals a high share of researchers who were not classified in the cores in both time periods (e.g. many researchers were classified in the periphery in the first and second periods). Further, many newcomers were classified in the semi-periphery or periphery in the second period. Similar pattern of many new researchers which were not connected to any previously existing authors was also found in other studies \cite{abbasi2012}.

Since the main interest of study is the stability of the cores of the obtained blockmodels, researchers not classified in the cores in at least one period can be removed from the visualization. Therefore, a new visualization can be presented in Figure~\ref{preh_b} consisting of two parts (one for each period) without the semi-periphery, periphery, newcomers, and departures. Instead, researchers classified in the cores in the first  but not in the second period are now called ``out-of-cores'' researchers and, similarly, researchers not classified in the cores in the first period but were classified in the core in the second period are now called ``into-cores'' researchers. Focusing on the core part of the sociology example, it can be observed that cores 1 and 2 merged in the second period, while core 3 splits into three cores in the second period. There are also many cores which disappear in the second period (out-of-cores researchers) and a lot of researchers not classified in the cores in the first but are classified in the cores in the second period. These into cores researchers usually join already the existing cores in the second period. 

\begin{figure}[!]
    \caption{Visualization of researchers' transitions in two time periods for sociology}
    \label{soc12_preh}
    \centering
    \begin{subfigure}[b]{0.45\textwidth}
    \centering
        \includegraphics[height=3.5cm]{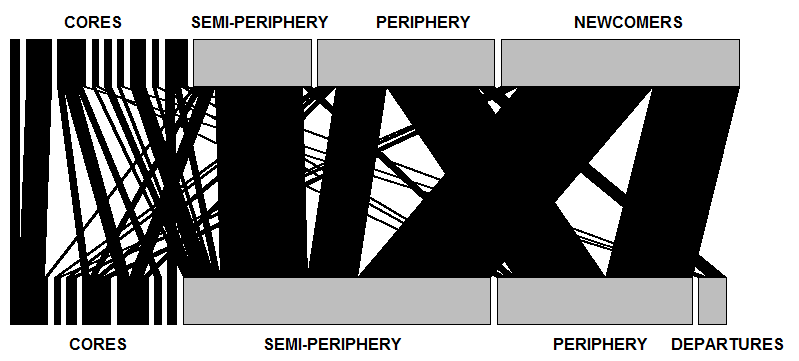}
        \caption{Transitions between the cores, semi-periphery and periphery}
        \label{preh_a}
    \end{subfigure}
  \hfill
    \begin{subfigure}[b]{0.45\textwidth}
    \centering
        \includegraphics[height=3.4cm]{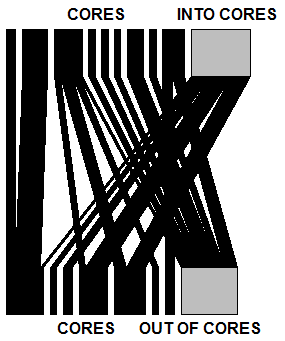}
        \caption{Transitionss between the cores, into cores and out of cores}
        \label{preh_b}
    \end{subfigure}
\end{figure}

Visualizations of researchers' transitions between the core, into cores and out of cores in the two periods are made for all analyzed scientific disciplines (Figure~\ref{all}). A relatively high share of into-cores and out-of-cores researchers in all analyzed scientific disciplines and some merging and splitting of cores in the core part of the visualized transitions can be seen. Here, the into-cores and out-of-cores researchers are seen as the primary source of instability of the core part of scientific disciplines. Although the share of into-cores researchers is higher than the share of out-of-cores researchers in almost all analyzed scientific disciplines, some scientific disciplines reveal the share of out-of-cores prevails over the share of into-cores researchers.

\begin{figure}
      \caption{Visualization of researchers' transitions between the cores in the two periods for all analyzed scientific disciplines (the black rectangles on the top and on the bottom correspond to the cores, the gray rectangles on the top correspond to the group of into cores while the gray rectangles on the bottom correspond to the group of out of cores)}
      \label{all}
    \includegraphics[width=\textwidth]{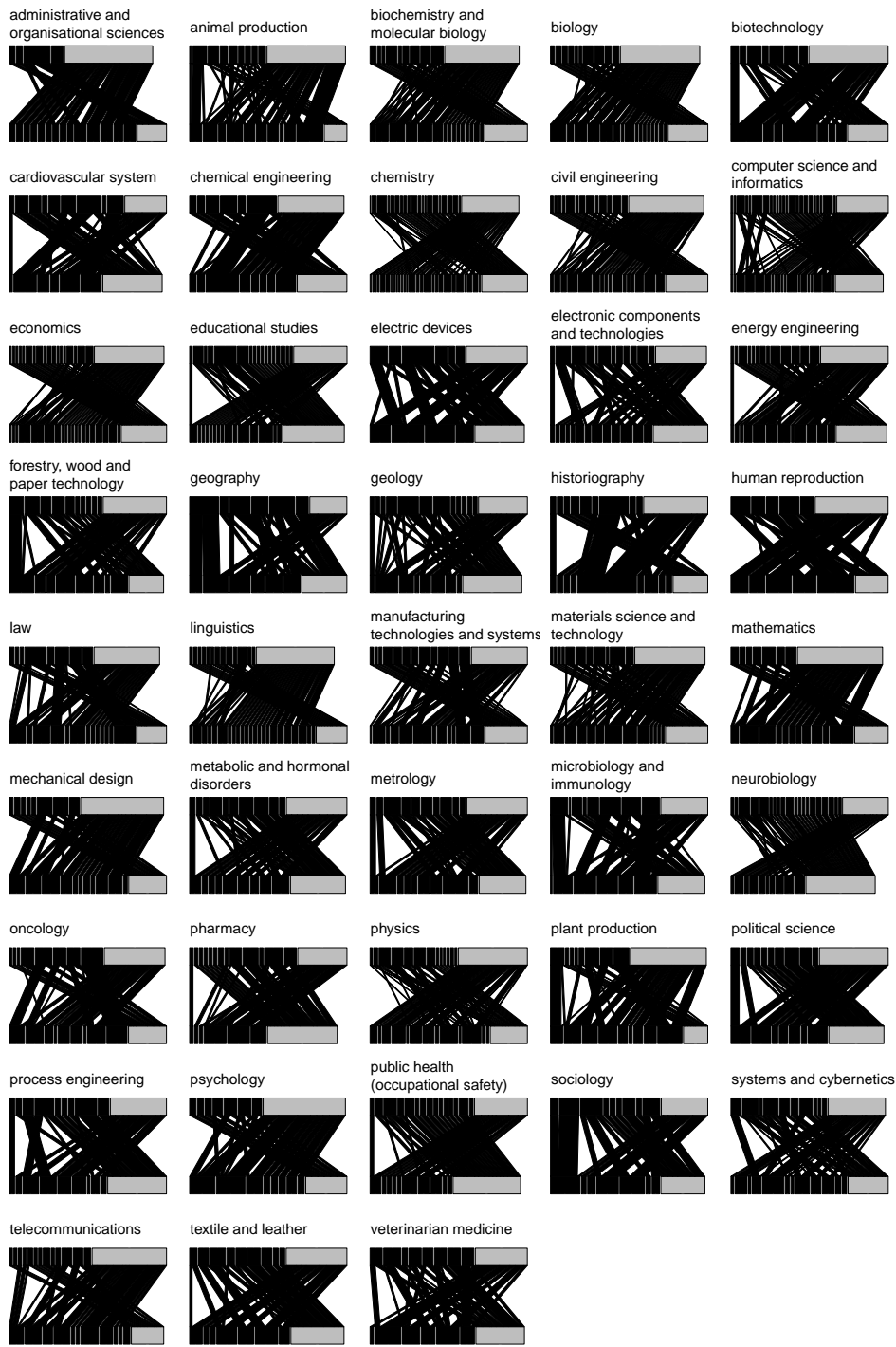}
\end{figure}

\begin{sidewaystable}
\caption{The considered scientific disciplines and their characteristics in the two periods -- part 1} 
\label{tabela_all}
\centering
\scriptsize
\linespread{0.5}
\begin{tabular}{|r|rrrrr|rrrrr|r|}
   \hline
                         & \multicolumn{5}{c|}{1991 - 2000}   & \multicolumn{5}{c|}{2001 - 2010}  &  \\ 
   Scientific discipline and scientific field & N & \begin{tabular}[c]{@{}l@{}} number \\ of cores \end{tabular} & semi-per. (\%) & per. (\%) & \begin{tabular}[c]{@{}l@{}} average \\ core size \end{tabular} & N & \begin{tabular}[c]{@{}l@{}} number \\ of cores \end{tabular} & semi-per. (\%) & per. (\%) & \begin{tabular}[c]{@{}l@{}} average \\ core size \end{tabular} &   \begin{tabular}[c]{@{}l@{}} \% of researchers \\ that left the cores \end{tabular}  \\ 
   \hline
\textbf{1. Natural sciences and mathematics} &  &  &  &  &  &  &  &  &  &  &  \\ 
   \hline
  biochemistry and molecular biology & 171 & 11 & 45.61 & 30.41 & 4.56 & 301 & 17 & 45.85 & 33.55 & 4.13 & 68.29 \\ 
  biology & 279 & 12 & 46.59 & 37.63 & 4.40 & 427 & 18 & 55.04 & 27.40 & 4.69 & 70.45 \\ 
  chemistry & 443 & 22 & 64.79 & 12.87 & 4.95 & 553 & 29 & 67.09 & 11.93 & 4.30 & 68.69 \\ 
  geology & 211 & 14 & 44.55 & 34.12 & 3.75 & 294 & 10 & 43.54 & 42.86 & 5.00 & 64.44 \\ 
  mathematics & 155 & 9 & 34.19 & 51.61 & 3.14 & 257 & 13 & 51.36 & 32.30 & 3.82 & 59.09 \\ 
  pharmacy & 228 & 15 & 51.75 & 25.00 & 4.08 & 285 & 8 & 71.58 & 14.04 & 6.83 & 77.36 \\ 
  physics & 307 & 15 & 61.89 & 15.96 & 5.23 & 451 & 15 & 65.63 & 12.42 & 7.62 & 52.94 \\ 
   \hline
\textbf{2. Engineering sciences and technologies} &  &  &  &  &  &  &  &  &  &  &  \\ 
   \hline
  chemical engineering & 153 & 8 & 60.13 & 24.18 & 4.00 & 197 & 10 & 66.50 & 18.78 & 3.62 & 62.50 \\ 
  civil engineering & 242 & 15 & 65.29 & 15.70 & 3.54 & 289 & 18 & 61.94 & 13.49 & 4.44 & 69.57 \\ 
  computer science and informatics & 401 & 25 & 51.87 & 27.43 & 3.61 & 565 & 21 & 63.36 & 21.24 & 4.58 & 62.65 \\ 
  electric devices & 70 & 8 & 31.43 & 25.71 & 5.00 & 97 & 6 & 38.14 & 26.80 & 8.50 & 56.67 \\ 
  electronic components and technologies & 116 & 14 & 36.21 & 23.28 & 3.92 & 135 & 15 & 40.74 & 25.93 & 3.46 & 61.70 \\ 
  energy engineering & 210 & 15 & 57.14 & 16.67 & 4.23 & 283 & 14 & 60.07 & 15.90 & 5.67 & 70.91 \\ 
  manufacturing technologies and systems & 217 & 14 & 45.62 & 28.57 & 4.67 & 276 & 14 & 52.90 & 23.55 & 5.42 & 50.00 \\ 
  materials science and technology & 253 & 13 & 60.08 & 17.39 & 5.18 & 332 & 15 & 62.05 & 13.86 & 6.15 & 61.40 \\ 
  mechanical design & 135 & 11 & 55.56 & 21.48 & 3.44 & 210 & 13 & 56.19 & 17.62 & 5.00 & 64.52 \\ 
  metrology & 74 & 12 & 39.19 & 20.27 & 3.00 & 100 & 11 & 35.00 & 30.00 & 3.89 & 56.67 \\ 
  process engineering & 92 & 11 & 52.17 & 11.96 & 3.67 & 93 & 11 & 48.39 & 17.20 & 3.56 & 66.67 \\ 
  systems and cybernetics & 240 & 12 & 54.17 & 21.67 & 5.80 & 291 & 14 & 58.42 & 17.53 & 5.83 & 48.28 \\ 
  telecommunications & 142 & 15 & 39.44 & 32.39 & 3.08 & 195 & 13 & 45.13 & 21.03 & 6.00 & 50.00 \\ 
  textile and leather & 100 & 11 & 55.00 & 12.00 & 3.67 & 131 & 9 & 61.07 & 11.45 & 5.14 & 63.64 \\ 
   \hline
\textbf{3. Medical sciences} &  &  &  &  &  &  &  &  &  &  &  \\ 
   \hline
  cardiovascular system & 224 & 11 & 54.91 & 16.52 & 7.11 & 268 & 8 & 67.54 & 12.69 & 8.83 & 57.81 \\ 
  human reproduction & 164 & 7 & 53.66 & 21.95 & 8.00 & 187 & 7 & 55.61 & 11.76 & 12.20 & 42.50 \\ 
  metabolic and hormonal disorders & 74 & 12 & 13.51 & 45.95 & 3.00 & 97 & 11 & 38.14 & 28.87 & 3.56 & 66.67 \\ 
  microbiology and immunology & 208 & 10 & 54.81 & 14.90 & 7.88 & 270 & 8 & 67.78 & 8.89 & 10.50 & 47.62 \\ 
  neurobiology & 296 & 22 & 51.01 & 25.68 & 3.45 & 366 & 12 & 69.67 & 16.67 & 5.00 & 81.16 \\ 
  oncology & 233 & 10 & 54.08 & 17.60 & 8.25 & 277 & 11 & 56.32 & 13.36 & 9.33 & 45.45 \\ 
  public health (occupational safety) & 195 & 17 & 34.87 & 41.54 & 3.07 & 264 & 12 & 56.06 & 29.55 & 3.80 & 80.43 \\ 
   \hline
  \end{tabular}
\end{sidewaystable}
 
\begin{sidewaystable}
\caption{The considered scientific disciplines and their characteristics in the two periods -- part 2} 
\label{tabela_all1}
\centering
\scriptsize
\linespread{0.5}
\begin{tabular}{|r|rrrrr|rrrrr|r|}
   \hline
                         & \multicolumn{5}{c|}{1991 - 2000}   & \multicolumn{5}{c|}{2001 - 2010}  &  \\ 
   Scientific discipline and scientific field & N & \begin{tabular}[c]{@{}l@{}} number \\ of cores \end{tabular} & semi-per. (\%) & per. (\%) & \begin{tabular}[c]{@{}l@{}} average \\ core size \end{tabular} & N & \begin{tabular}[c]{@{}l@{}} number \\ of cores \end{tabular} & semi-per. (\%) & per. (\%) & \begin{tabular}[c]{@{}l@{}} average \\ core size \end{tabular} &   \begin{tabular}[c]{@{}l@{}} \% of researchers \\ that left the cores \end{tabular}  \\ 
   \hline  
\textbf{4. Biotechnical sciences} &  &  &  &  &  &  &  &  &  &  &  \\ 
   \hline
  animal production & 127 & 11 & 64.57 & 6.30 & 4.11 & 158 & 12 & 48.73 & 7.59 & 6.90 & 35.14 \\ 
  biotechnology & 111 & 9 & 49.55 & 18.92 & 5.00 & 196 & 9 & 58.16 & 13.78 & 7.86 & 60.00 \\ 
  forestry, wood and paper technology & 149 & 12 & 62.42 & 11.41 & 3.90 & 153 & 10 & 53.59 & 11.76 & 6.62 & 43.59 \\ 
  plant production & 247 & 10 & 68.83 & 12.96 & 5.62 & 312 & 11 & 66.35 & 8.33 & 8.78 & 35.56 \\ 
  veterinarian medicine & 171 & 10 & 60.23 & 8.19 & 6.75 & 168 & 8 & 61.31 & 5.36 & 9.33 & 51.85 \\ 
   \hline
\textbf{5. Social sciences} &  &  &  &  &  &  &  &  &  &  &  \\ 
   \hline
  administrative and organisational sciences & 96 & 6& 17.71 & 64.58 & 4.25 & 155 & 16 & 34.19 & 41.94 & 2.64 & 58.82 \\ 
  economics & 338 & 20 & 49.11 & 32.84 & 3.39 & 477 & 22 & 61.01 & 21.17 & 4.25 & 73.77 \\ 
  educational studies & 303 & 19 & 38.28 & 39.60 & 3.94 & 404 & 17 & 51.24 & 33.91 & 4.00 & 76.12 \\ 
  law & 157 & 10 & 24.20 & 47.13 & 5.62 & 204 & 15 & 31.37 & 35.78 & 5.15 & 40.00 \\ 
  political science & 98 & 12 & 23.47 & 48.98 & 2.70 & 142 & 9 & 45.77 & 31.69 & 4.57 & 74.07 \\ 
  psychology & 92 & 9 & 29.35 & 48.91 & 2.86 & 131 & 10 & 38.17 & 36.64 & 4.12 & 65.00 \\ 
  sociology & 135 & 11 & 28.15 & 42.22 & 4.44 & 203 & 9 & 48.77 & 31.03 & 5.86 & 45.00 \\ 
   \hline
\textbf{6. Humanities} &  &  &  &  &  &  &  &  &  &  &  \\ 
   \hline
  geography & 87 & 8 & 33.33 & 35.63 & 4.50 & 117 & 8 & 53.85 & 24.79 & 4.17 & 40.74 \\ 
  historiography & 186 & 11 & 9.14 & 74.73 & 3.33 & 251 & 11 & 23.11 & 60.56 & 4.56 & 43.33 \\ 
  linguistics & 196 & 15 & 10.71 & 71.94 & 2.62 & 296 & 25 & 22.64 & 55.07 & 2.87 & 61.76 \\ 
  \hline
\end{tabular}
\end{sidewaystable}

The number and the size of the cores, the size of the semi-periphery and the size of the periphery vary across scientific disciplines (see Figure~\ref{trend} and Table~\ref{tabela_all} and \ref{tabela_all1}). For example, the discipline administrative and organizational sciences consist of 6 cores in the first period and 16 cores in the second. As it is shown in the latter sections, most of the existing cores in the second time period emerged from the non-core part of the network. The newly emerged cores are of smaller size (in average 2.64 researchers) than the cores in the first time period. These observations indicate that the scientific collaboration might become seen as more beneficial by the researchers from the field administrative and organizational sciences. Keep in mind that this is a scientific discipline with a relatively low number of researchers (96 researchers in the first and 155 researchers - with at least one published scientific bibliographic unit - in the second time period).

There are usually higher number of cores in the disciplines with a higher number of researchers which is expected since the personal limits of each researcher to cooperate with a limited number of coauthors and produce a limited number of publications \cite{kronegger2011, de1986little}. One such example is physics with 307 researches (with at least one published scientific bibliographic unit) in the first and 451 researchers in the second period. There are 15 cores revealed in both the first and second period. As this is the case for almost all scientific disciplines, the average core size is higher in the second period than in the first. The decrease in the average core size is usually the consequence of many cores of size two emerged (e.q., biochemistry and molecular biology, chemistry, law). These can consist of any kind of researches, for example, a core of size three can consist of a student and his mentor. However, it is assumed that any kind of scientific collaboration of which the output is scientific bibliographic unit, requires very intensive collaboration - exchange of knowledge and ideas. The pairs of scientists, collaborating as researchers are also very common in the field of social network analysis. One example could be Borgatti and Everett.

However, some studies found that type (e.g., natural vs. social or office vs. lab or theoretical vs. empirical) of a scientific discipline affect the size of a research teams \cite{melin2000, kyvik2003, hu2014}. Here, the highest average core size in the first period is observed in oncology (8.3 researchers) and human reproduction (8.0 researchers), while the lowest average core size in the first period is observed in linguistics (2.6 researchers) and psychology (2.9 researchers). In general, the overall average number of cores is similar in both periods (around 11 cores), while the overall average size of the cores is increasing in time (from 4.4 to 5.6 researchers, $p<0.01$), as confirmed by Amat and Perruchas \cite{amat2015}.

Following the distinction between the natural and technical sciences on one side, and the social sciences and humanities on the other, it can be concluded that the average size of the cores is increasing, especially in the natural and technical sciences (form 4.6 to 6.1 researchers, $p<0.01$), while in the social sciences and humanities, the growth of the average size of cores (from 3.8 to 4.2 researchers, $p=0.30$) is not statistically significant. In general, the average core size is lower in the social sciences and humanities in both periods (for 0.95 researchers in the first and 1.85 researchers in the second time period; $p<0.05$ and $p<0.01$ susbequently) (Figure~\ref{trend}).

\begin{figure}[!]
    \centering
      \caption{The average core size and the average size of the periphery by field and time period}
      \label{trend}
      \includegraphics[width=1\textwidth]{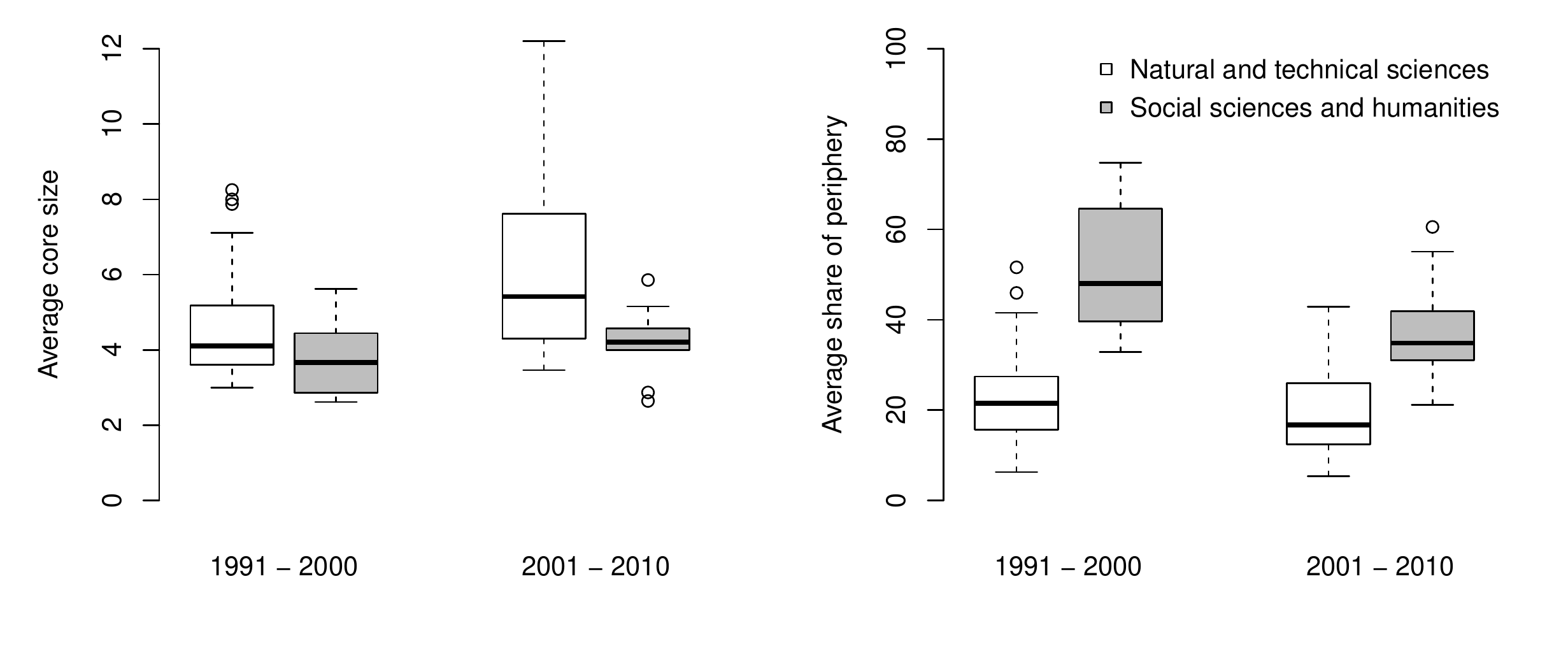}
\end{figure}

Solo authors or authors who published only in co-authorship with authors from outside the discipline are classified in the periphery. The average share of these authors among analyzed scientific disciplines is decreasing in time (from 29~\% to 23~\% average share of the periphery, $p=0.06$). The biggest reduction in the percentage of the periphery in the second period is observed in criminology and social work (a 65~\% decrease). In some scientific disciplines, the percentage of the periphery increased in the second period. These are mainly disciplines from the natural and technical fields. However, the size of the periphery is greater in fields of the social sciences and humanities (the average size is 44~\%) than in scientific disciplines classified in the natural and technical sciences (the average size is 21~\%)($p<0.01$). In addition, the average share of the periphery decreases from the first to the second period espcially in the natural and technical sciences (from 51 to 37~\%, $p<0.05$) while the difference in the average share of periphery is not statistically significant ($p=0.11$) in the social sciences and humanities (the share of the periphery decreased from 23 to 19~\%) (Figure~\ref{trend}).

\section{Stability of the obtained blockmodel structures}

In this section, the stability of cores is studied according to different operationalisations of core stability. Although the presented visualizations of researchers' transitions between two time periods (Figure~\ref{all}) are a very efficient tool for studying the stability of the cores obtained but whose interpretation is complex, the values of the indices proposed in Section~\ref{indices} are calculated. These indices are more objective operationalizations of core stability and allow us to compare the values calculated for different scientific disciplines. The scientific disciplines are then clustered according to the calculated indices. The groups of scientific disciplines thus obtained are further analyzed.

In the second part, the operationalization of the stability of cores is restricted to one of the described indices for measuring core stability, namely, as applied in Cugmas et al. \cite{cugmas2015} only the splitting of cores and the out-of-cores researchers are seen as features indicating lower stability of the cores. The  differences in the mean stability of cores among different scientific fields are studied using linear regression. Some further controlling explanatory variables are also included in the model.

First, the values of each presented index for each analyzed scientific discipline are shown in Table~\ref{all_indices} and provide the basis for all further analyses. In this table, one sees that the values of the Adjusted Rand Index and the adjusted Wallace indices are relatively large, while the others are relatively small. This is due to the high share of into-cores and out-of-cores researchers which are not considered when calculating the values of the Adjusted Rand Index and the adjusted Wallace indices for each scientific discipline. The high values of the first three indices and the low values of the others confirm that the into-cores researchers and out-of-cores researchers are the biggest source of the obtained cores' instability.

\begin{table}[!]
\caption{The values of different indices for measuring the stability of cores for all analyzed scientific disciplines by obtained clusters}
\label{all_indices}
\footnotesize
\centering
\begin{tabular}{*{10}{l|ccc|ccc|ccc}}
\hline
Discipline & \rot{90}{ARI} & \rot{90}{AWI'} & \rot{90}{AWI''} & \rot{90}{MARI1} & \rot{90}{MAWIS1} & \rot{90}{MAWIM1} & \rot{90}{MARI2} & \rot{90}{MAWIS2} & \rot{90}{MAWIM2} \\ \hline
\multicolumn{10}{l}{\textbf{Cluster 1 (unstable)}}   \\ 
     biochemistry and molecular biology & \cellcolor{gray!16}-0.16& \cellcolor{gray!11}-0.11& \cellcolor{gray!27}-0.27& \cellcolor{gray!2}-0.02& \cellcolor{gray!0}0.00& \cellcolor{gray!0}0.00& \cellcolor{gray!3}-0.03& \cellcolor{gray!0}0.00& \cellcolor{gray!0}0.00\\
   geology & \cellcolor{gray!9}0.09& \cellcolor{gray!11}0.11& \cellcolor{gray!7}0.07& \cellcolor{gray!0}0.00& \cellcolor{gray!2}0.02& \cellcolor{gray!0}0.00& \cellcolor{gray!0}0.00& \cellcolor{gray!1}0.01& \cellcolor{gray!3}0.03\\
  psychology & \cellcolor{gray!4}0.04& \cellcolor{gray!7}0.07& \cellcolor{gray!3}0.03& \cellcolor{gray!0}0.00& \cellcolor{gray!2}0.02& \cellcolor{gray!0}0.00& \cellcolor{gray!9}-0.09& \cellcolor{gray!0}0.00& \cellcolor{gray!1}0.01\\
  pharmacy& \cellcolor{gray!50}0.50& \cellcolor{gray!69}0.69& \cellcolor{gray!39}0.39& \cellcolor{gray!1}0.01& \cellcolor{gray!3}0.03& \cellcolor{gray!1}0.01& \cellcolor{gray!9}0.09& \cellcolor{gray!1}0.01& \cellcolor{gray!3}0.03\\
  physics& \cellcolor{gray!40}0.40& \cellcolor{gray!67}0.67& \cellcolor{gray!28}0.28& \cellcolor{gray!3}0.03& \cellcolor{gray!15}0.15& \cellcolor{gray!4}0.04& \cellcolor{gray!3}0.03& \cellcolor{gray!2}0.02& \cellcolor{gray!8}0.08\\
  neurobiology & \cellcolor{gray!43}0.43& \cellcolor{gray!68}0.68& \cellcolor{gray!32}0.32& \cellcolor{gray!0}0.00& \cellcolor{gray!8}0.08& \cellcolor{gray!1}0.01& \cellcolor{gray!3}0.03& \cellcolor{gray!1}0.01& \cellcolor{gray!9}0.09\\
  materials science and technology & \cellcolor{gray!33}0.33& \cellcolor{gray!26}0.26& \cellcolor{gray!45}0.45& \cellcolor{gray!1}0.01& \cellcolor{gray!6}0.06& \cellcolor{gray!2}0.02& \cellcolor{gray!2}0.02& \cellcolor{gray!1}0.01& \cellcolor{gray!4}0.04\\
  public health (occupational safety)  & \cellcolor{gray!37}0.37& \cellcolor{gray!32}0.32& \cellcolor{gray!42}0.42& \cellcolor{gray!0}0.00& \cellcolor{gray!3}0.03& \cellcolor{gray!0}0.00& \cellcolor{gray!0}0.00& \cellcolor{gray!0}0.00& \cellcolor{gray!2}0.02\\
  biology & \cellcolor{gray!38}0.38& \cellcolor{gray!48}0.48& \cellcolor{gray!31}0.31& \cellcolor{gray!1}0.01& \cellcolor{gray!9}0.09& \cellcolor{gray!2}0.02& \cellcolor{gray!2}-0.02& \cellcolor{gray!0}0.00& \cellcolor{gray!4}0.04\\
  educational studies & \cellcolor{gray!32}0.32& \cellcolor{gray!34}0.34& \cellcolor{gray!31}0.31& \cellcolor{gray!0}0.00& \cellcolor{gray!4}0.04& \cellcolor{gray!1}0.01& \cellcolor{gray!2}-0.02& \cellcolor{gray!1}0.01& \cellcolor{gray!7}0.07\\
  linguistics& \cellcolor{gray!43}0.43& \cellcolor{gray!29}0.29& \cellcolor{gray!82}0.82& \cellcolor{gray!0}0.00& \cellcolor{gray!16}0.16& \cellcolor{gray!3}0.03& \cellcolor{gray!3}0.03& \cellcolor{gray!0}0.00& \cellcolor{gray!8}0.08\\
  electric devices & \cellcolor{gray!50}0.50& \cellcolor{gray!46}0.46& \cellcolor{gray!56}0.56& \cellcolor{gray!3}0.03& \cellcolor{gray!11}0.11& \cellcolor{gray!6}0.06& \cellcolor{gray!4}0.04& \cellcolor{gray!3}0.03& \cellcolor{gray!7}0.07\\
  metrology & \cellcolor{gray!42}0.42& \cellcolor{gray!44}0.44& \cellcolor{gray!40}0.40& \cellcolor{gray!1}0.01& \cellcolor{gray!16}0.16& \cellcolor{gray!4}0.04& \cellcolor{gray!0}0.00& \cellcolor{gray!2}0.02& \cellcolor{gray!9}0.09\\
  \hline
  Standardized means &  0.31 & 0.36 & 0.31 & 0.01 & 0.07 & 0.07 & 0.01 & 0.01 & 0.05 \\ 
  \hline \hline
  \multicolumn{10}{l}{\textbf{Cluster 2 (average)}}   \\ 
  mathematics & \cellcolor{gray!100}1.00& \cellcolor{gray!100}1.00& \cellcolor{gray!100}1.00& \cellcolor{gray!5}0.05& \cellcolor{gray!27}0.27& \cellcolor{gray!8}0.08& \cellcolor{gray!1}0.01& \cellcolor{gray!1}0.01& \cellcolor{gray!9}0.09 \\
  civil engineering & \cellcolor{gray!86}0.86& \cellcolor{gray!76}0.76& \cellcolor{gray!100}1.00& \cellcolor{gray!1}0.01& \cellcolor{gray!14}0.14& \cellcolor{gray!2}0.02& \cellcolor{gray!1}0.01& \cellcolor{gray!1}0.01& \cellcolor{gray!6}0.06\\
  energy engineering  & \cellcolor{gray!81}0.81& \cellcolor{gray!75}0.75& \cellcolor{gray!88}0.88& \cellcolor{gray!1}0.01& \cellcolor{gray!15}0.15& \cellcolor{gray!2}0.02& \cellcolor{gray!5}0.05& \cellcolor{gray!1}0.01& \cellcolor{gray!9}0.09\\
  systems and cybernetics & \cellcolor{gray!70}0.70& \cellcolor{gray!57}0.57& \cellcolor{gray!92}0.92& \cellcolor{gray!4}0.04& \cellcolor{gray!19}0.19& \cellcolor{gray!9}0.09& \cellcolor{gray!12}0.12& \cellcolor{gray!4}0.04& \cellcolor{gray!15}0.15\\
  computer science and informatics & \cellcolor{gray!63}0.63& \cellcolor{gray!57}0.57& \cellcolor{gray!71}0.71& \cellcolor{gray!1}0.01& \cellcolor{gray!19}0.19& \cellcolor{gray!3}0.03& \cellcolor{gray!4}0.04& \cellcolor{gray!2}0.02& \cellcolor{gray!14}0.14\\
  telecommunications& \cellcolor{gray!69}0.69& \cellcolor{gray!89}0.89& \cellcolor{gray!56}0.56& \cellcolor{gray!4}0.04& \cellcolor{gray!35}0.35& \cellcolor{gray!8}0.08& \cellcolor{gray!6}0.06& \cellcolor{gray!2}0.02& \cellcolor{gray!9}0.09\\
  electronic components and technologies & \cellcolor{gray!62}0.62& \cellcolor{gray!47}0.47& \cellcolor{gray!91}0.91& \cellcolor{gray!1}0.01& \cellcolor{gray!13}0.13& \cellcolor{gray!3}0.03& \cellcolor{gray!11}0.11& \cellcolor{gray!3}0.03& \cellcolor{gray!19}0.19\\
  mechanical design & \cellcolor{gray!100}1.00& \cellcolor{gray!100}1.00& \cellcolor{gray!100}1.00& \cellcolor{gray!4}0.04& \cellcolor{gray!23}0.23& \cellcolor{gray!6}0.06& \cellcolor{gray!4}0.04& \cellcolor{gray!1}0.01& \cellcolor{gray!9}0.09\\
  process engineering & \cellcolor{gray!84}0.84& \cellcolor{gray!72}0.72& \cellcolor{gray!100}1.00& \cellcolor{gray!2}0.02& \cellcolor{gray!14}0.14& \cellcolor{gray!4}0.04& \cellcolor{gray!10}0.1& \cellcolor{gray!4}0.04& \cellcolor{gray!17}0.17\\
  textile and leather & \cellcolor{gray!80}0.80& \cellcolor{gray!80}0.80& \cellcolor{gray!80}0.80& \cellcolor{gray!3}0.03& \cellcolor{gray!18}0.18& \cellcolor{gray!4}0.04& \cellcolor{gray!0}0.00& \cellcolor{gray!3}0.03& \cellcolor{gray!10}0.10\\
  human reproduction & \cellcolor{gray!40}0.40& \cellcolor{gray!93}0.93& \cellcolor{gray!26}0.26& \cellcolor{gray!8}0.08& \cellcolor{gray!28}0.28& \cellcolor{gray!14}0.14& \cellcolor{gray!12}0.12& \cellcolor{gray!6}0.06& \cellcolor{gray!14}0.14\\
  metabolic and hormonal disorders & \cellcolor{gray!73}0.73& \cellcolor{gray!100}1.00& \cellcolor{gray!57}0.57& \cellcolor{gray!2}0.02& \cellcolor{gray!7}0.07& \cellcolor{gray!1}0.01& \cellcolor{gray!2}-0.02& \cellcolor{gray!1}0.01& \cellcolor{gray!6}0.06\\
  chemistry & \cellcolor{gray!60}0.60& \cellcolor{gray!46}0.46& \cellcolor{gray!89}0.89& \cellcolor{gray!1}0.01& \cellcolor{gray!17}0.17& \cellcolor{gray!2}0.02& \cellcolor{gray!4}0.04& \cellcolor{gray!1}0.01& \cellcolor{gray!17}0.17\\
  forestry, wood and paper technology & \cellcolor{gray!64}0.64& \cellcolor{gray!69}0.69& \cellcolor{gray!60}0.60& \cellcolor{gray!6}0.06& \cellcolor{gray!34}0.34& \cellcolor{gray!14}0.14& \cellcolor{gray!10}0.10& \cellcolor{gray!5}0.05& \cellcolor{gray!15}0.15\\
  animal production  & \cellcolor{gray!49}0.49& \cellcolor{gray!51}0.51& \cellcolor{gray!47}0.47& \cellcolor{gray!6}0.06& \cellcolor{gray!19}0.19& \cellcolor{gray!13}0.13& \cellcolor{gray!6}0.06& \cellcolor{gray!1}0.01& \cellcolor{gray!6}0.06\\
  veterinarian medicine & \cellcolor{gray!52}0.52& \cellcolor{gray!68}0.68& \cellcolor{gray!43}0.43& \cellcolor{gray!4}0.04& \cellcolor{gray!15}0.15& \cellcolor{gray!5}0.05& \cellcolor{gray!13}0.13& \cellcolor{gray!5}0.05& \cellcolor{gray!9}0.09\\
  biotechnology & \cellcolor{gray!73}0.73& \cellcolor{gray!100}1.00& \cellcolor{gray!57}0.57& \cellcolor{gray!4}0.04& \cellcolor{gray!14}0.14& \cellcolor{gray!5}0.05& \cellcolor{gray!1}-0.01& \cellcolor{gray!1}0.01& \cellcolor{gray!4}0.04\\
  economics & \cellcolor{gray!71}0.71& \cellcolor{gray!64}0.64& \cellcolor{gray!80}0.80& \cellcolor{gray!1}0.01& \cellcolor{gray!14}0.14& \cellcolor{gray!1}0.01& \cellcolor{gray!1}0.01& \cellcolor{gray!1}0.01& \cellcolor{gray!7}0.07\\
  administrative and organisational sciences & \cellcolor{gray!80}0.80& \cellcolor{gray!67}0.67& \cellcolor{gray!100}1.00& \cellcolor{gray!6}0.06& \cellcolor{gray!11}0.11& \cellcolor{gray!9}0.09& \cellcolor{gray!5}0.05& \cellcolor{gray!1}0.01& \cellcolor{gray!19}0.19\\
  law & \cellcolor{gray!58}0.58& \cellcolor{gray!80}0.80& \cellcolor{gray!45}0.45& \cellcolor{gray!9}0.09& \cellcolor{gray!29}0.29& \cellcolor{gray!17}0.17& \cellcolor{gray!14}-0.14& \cellcolor{gray!6}0.06& \cellcolor{gray!22}0.22\\
  political science & \cellcolor{gray!86}0.86& \cellcolor{gray!100}1.00& \cellcolor{gray!75}0.75& \cellcolor{gray!1}0.01& \cellcolor{gray!13}0.13& \cellcolor{gray!2}0.02& \cellcolor{gray!3}-0.03& \cellcolor{gray!1}0.01& \cellcolor{gray!5}0.05\\
  historiography & \cellcolor{gray!57}0.57& \cellcolor{gray!68}0.68& \cellcolor{gray!49}0.49& \cellcolor{gray!8}0.08& \cellcolor{gray!39}0.39& \cellcolor{gray!20}0.20& \cellcolor{gray!5}0.05& \cellcolor{gray!7}0.07& \cellcolor{gray!9}0.09\\
  \hline
  Standardized means & 0.71 & 0.75 &  0.73 &  0.04 &  0.20 & 0.17  & 0.04 & 0.03  &  0.12\\ 
  \hline \hline
    \multicolumn{10}{l}{\textbf{Cluster 3 (stable)}}   \\ 
  plant production & \cellcolor{gray!90}0.90& \cellcolor{gray!84}0.84& \cellcolor{gray!97}0.97& \cellcolor{gray!15}0.15& \cellcolor{gray!45}0.45& \cellcolor{gray!34}0.34& \cellcolor{gray!11}0.11& \cellcolor{gray!5}0.05& \cellcolor{gray!19}0.19\\
  oncology & \cellcolor{gray!89}0.89& \cellcolor{gray!85}0.85& \cellcolor{gray!93}0.93& \cellcolor{gray!11}0.11& \cellcolor{gray!42}0.42& \cellcolor{gray!23}0.23& \cellcolor{gray!12}0.12& \cellcolor{gray!11}0.11& \cellcolor{gray!35}0.35\\
  chemical engineering & \cellcolor{gray!100}1.00& \cellcolor{gray!100}1.00& \cellcolor{gray!100}1.00& \cellcolor{gray!8}0.08& \cellcolor{gray!33}0.33& \cellcolor{gray!14}0.14& \cellcolor{gray!9}0.09& \cellcolor{gray!8}0.08& \cellcolor{gray!30}0.30\\
  manufacturing technologies and systems & \cellcolor{gray!95}0.95& \cellcolor{gray!90}0.90& \cellcolor{gray!100}1.00& \cellcolor{gray!6}0.06& \cellcolor{gray!43}0.43& \cellcolor{gray!12}0.12& \cellcolor{gray!11}0.11& \cellcolor{gray!7}0.07& \cellcolor{gray!25}0.25\\
  microbiology and immunology & \cellcolor{gray!88}0.88& \cellcolor{gray!86}0.86& \cellcolor{gray!91}0.91& \cellcolor{gray!9}0.09& \cellcolor{gray!32}0.32& \cellcolor{gray!16}0.16& \cellcolor{gray!25}0.25& \cellcolor{gray!16}0.16& \cellcolor{gray!26}0.26\\
  cardiovascular system & \cellcolor{gray!100}1.00& \cellcolor{gray!100}1.00& \cellcolor{gray!100}1.00& \cellcolor{gray!6}0.06& \cellcolor{gray!30}0.30& \cellcolor{gray!10}0.10& \cellcolor{gray!32}0.32& \cellcolor{gray!18}0.18& \cellcolor{gray!30}0.30\\
  sociology & \cellcolor{gray!52}0.52& \cellcolor{gray!55}0.55& \cellcolor{gray!50}0.50& \cellcolor{gray!6}0.06& \cellcolor{gray!36}0.36& \cellcolor{gray!16}0.16& \cellcolor{gray!25}0.25& \cellcolor{gray!14}0.14& \cellcolor{gray!23}0.23\\
  geography& \cellcolor{gray!36}0.36& \cellcolor{gray!29}0.29& \cellcolor{gray!49}0.49& \cellcolor{gray!3}0.03& \cellcolor{gray!15}0.15& \cellcolor{gray!12}0.12& \cellcolor{gray!22}0.22& \cellcolor{gray!12}0.12& \cellcolor{gray!21}0.21\\
  \hline
  Standardized means & 0.81  & 0.79  & 0.85  &  0.08& 0.35 &0.02  &0.18  &0.11  &  0.26\\ 
  \hline \hline
  \end{tabular}
  \end{table}

\subsection{Clustering of scientific disciplines according to different operationalisations of the stability of cores}

Based on the calculated standardzied indices (see Table~\ref{all_indices} for non-standardized values of the indices), the analyzed scientific disciplines are clustered using Ward's agglomerative clustering method and squared Euclidean distance. Using the GAP Statistics \citep{tibshirani2001} and the obtained dendrogram three clusters were chosen. By observing the means of the calculated indices for each cluster (see Table~\ref{all_indices}), the obtained clusters can be ordered from the least stable (Cluster 1) to the most stable cluster (Cluster 3). Cluster 2 is named average since the values of all the indices are closest to the global means, compared to the other groups. Table~\ref{basic_des} summarizes some descriptive statistics of other blockmodels' characteristics:

\begin{itemize}
  \item \emph{The percentage of the into-cores (\% into-cores) and out-of-cores (\% out-of-cores) researchers.} The percentage of into-cores researchers is defined as the ratio between the number of researchers not in the cores in the first period and the number of researchers classified in the cores in the first period. On the other hand, the percentage of out-of-cores researchers is defined as the ratio between the number of researchers who joined the cores in the second period and the number of researchers classified in cores in the second period. Since Slovenian scientific disciplines are generally growing, the average share of into-cores researchers is lower than the share of out-of-cores researchers. However, a higher percentage of into-cores than out-of-cores researchers is typical for the unstable cluster of scientific disciplines.
  \item \emph{The overall average core size (core size) and the overall number of researchers across clusters of scientific disciplines (\# of res.)}. The average size of the cores is relatively small, the smallest is in the case of an unstable cluster (3.9 researchers) and the highest in the case of the most stable cluster (5.8 researchers). While a higher average core size is typical for more stable scientific disciplines, a higher number of researchers per discipline is related to less stable scientific disciplines.
  \item \emph{The number of scientific disciplines.} The average cluster, according to the values of the stability measures, has the highest number of scientific disciplines, followed by the unstable and the stable cluster. 
\end{itemize}

In the Slovenian Research Agency's classification scheme, the scientific fields are further divided into several scientific disciplines and then into scientific sub-disciplines. Based on this, most scientific disciplines from the fields of engineering sciences and technologies (9 out of 14), biotechnological sciences (4 out of 5) and social sciences (4 out of 7) were classified in the unstable cluster. Most (5 out of 7) scientific disciplines from the natural sciences and mathematics were classified in the average cluster and three out of seven scientific disciplines from the field of medical sciences were classified in the most stable cluster. We can say the most stable scientific disciplines are from medical sciences and the most unstable from the technical field and social sciences. Similarly, Melin \cite{melin2000} concluded that researchers from the medical sciences field almost always work in teams and from time to time collaborate with other teams. Kyvik \cite{kyvik2003} reports that the greatest number of multi-authored papers in Norway is in medicine.

\begin{table}[!]
\caption{Basic descriptive statistics of the obtained clusters (averages on the level of clusters are reported)}
\label{basic_des}
\footnotesize
\centering
\begin{tabular}{|l|cccc|} \hline
Cluster                              & \% into cores &  \% out of cores & core size & \# of res. \\  \hline
Cluster 1 (N = 13) (unstable)            & 72 & 67 & 3.9 & 322 \\
Cluster 2 (N = 22) (average)         & 69 & 58 & 4.2 & 274 \\ 
Cluster 3 (N = 8)~ (stable)           & 53 & 48 & 5.8 & 272  \\  \hline
\end{tabular}
\end{table}

Since a scientific discipline's affiliation with a certain cluster is a categorical variable, one can check if the basic characteristics presented in Table~\ref{basic_des} can be used to predict the cluster in which a given scientific discipline belongs. To do this, discriminant analysis can be used. Since there are three clusters of scientific disciplines, two discriminant functions can be calculated based on the four explanatory variables presented in Table~\ref{basic_des}. Only the first discriminant function is statistically significant ($p<0.01$), meaning that based on the four explanatory variables one can separate well between the stable cluster (cluster 3) on one side and average and unstable clusters (clusters 1 and 2) on the other. The discriminant functions are defined as linear combinations of explanatory variables. In Figure~\ref{disk}, the first discriminant function is visualized. Here, the highest values of the first discriminant function are characterized by higher mean percentage of into-cores (0.74) and percentage out-of-cores researchers (0.20) and a lower average number of researchers in the cores (-0.31). The value of the standardized canonical coefficient of the explanatory variable 'number of researchers' is relatively low (-0.09) and is therefore not shown in Figure~\ref{disk}. The centroids for each cluster are also marked along with the distribution of the standardized discriminant function for the disciplines by clusters.  

\begin{figure}[!]
    \caption{The distribution of standardized values of the first canonical discriminant function by clusters}
    \label{disk}
    \centering
        \includegraphics[width=\textwidth]{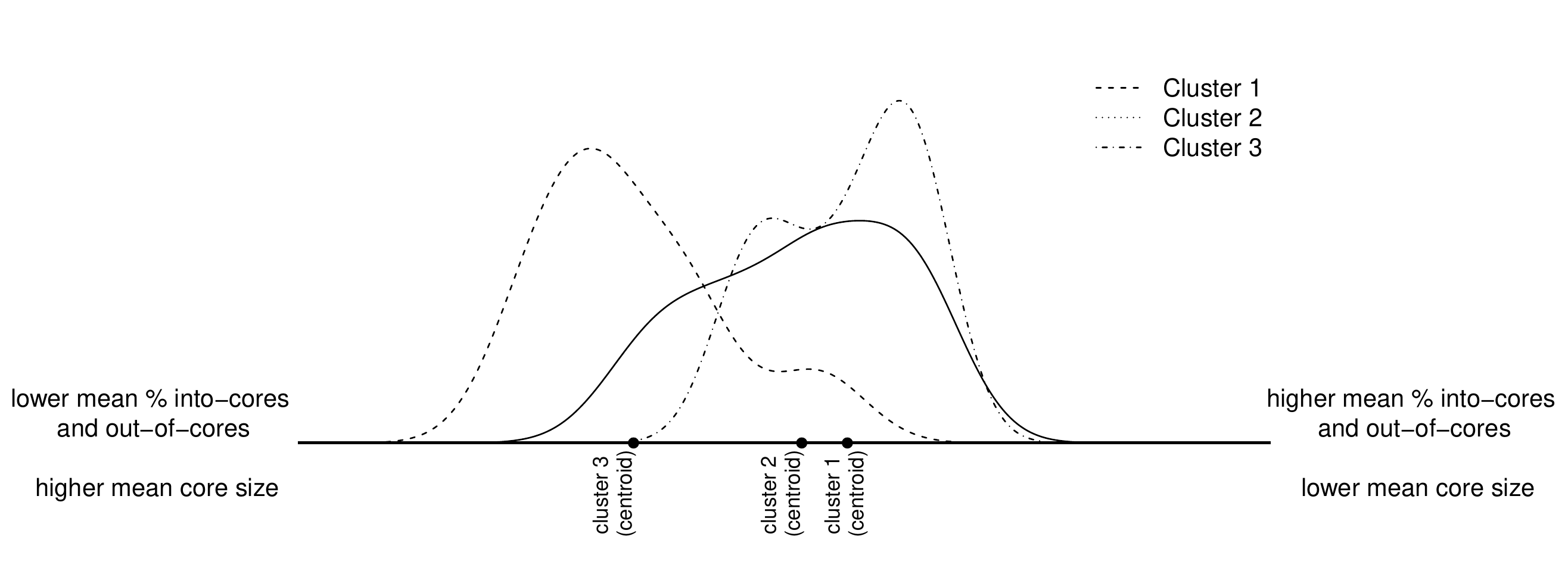}
\end{figure}

From each cluster of the scientific disciplines one was chosen to represent the cluster (the closest one to the centroid). The representative of the unstable cluster is the scientific discipline educational studies. Here, many into-cores and out-of-cores researchers can be seen. Most pairs of researchers classified in the same core at the first time point were not classified in the same core in the second period. The representative of the average cluster is the scientific discipline of textile and leather. Here, the share of out-of-cores and into-cores researchers is lower. Some relatively large cores which remain relatively stable in the second period can also be observed. This is more typical for the representative of the stable cluster, namely microbiology and immunology. 

\begin{figure}[H]
    \caption{Visualizations of researchers' transitions between the cores into cores and out of cores for the two periods for the scientific discipline closest to the centroid in each cluster (the black rectangles on the top and on the bottom correspond to the cores, the gray rectangles on the top correspond to the group of into cores while the gray rectangles on the bottom correspond to the group of out of cores)}
    \label{examples}
        \centering
    \begin{subfigure}[b]{0.30\textwidth}
        \includegraphics[width=\textwidth]{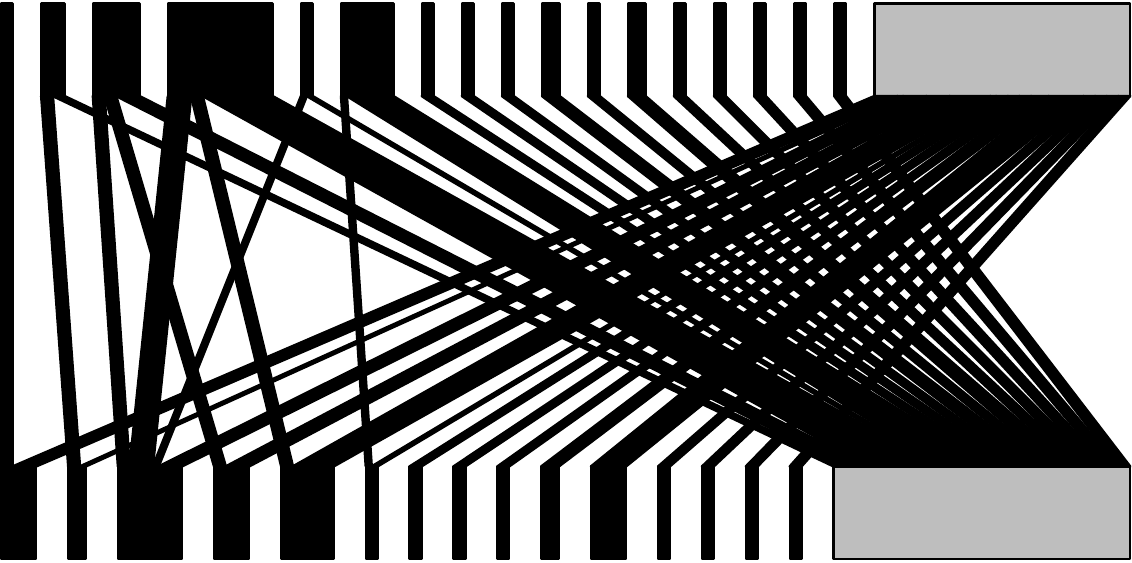}
        \caption{educational studies \\ (N=379) ~ \\}
        \label{educ}
    \end{subfigure}
    ~ \hfill
    \begin{subfigure}[b]{0.30\textwidth}
        \includegraphics[width=\textwidth]{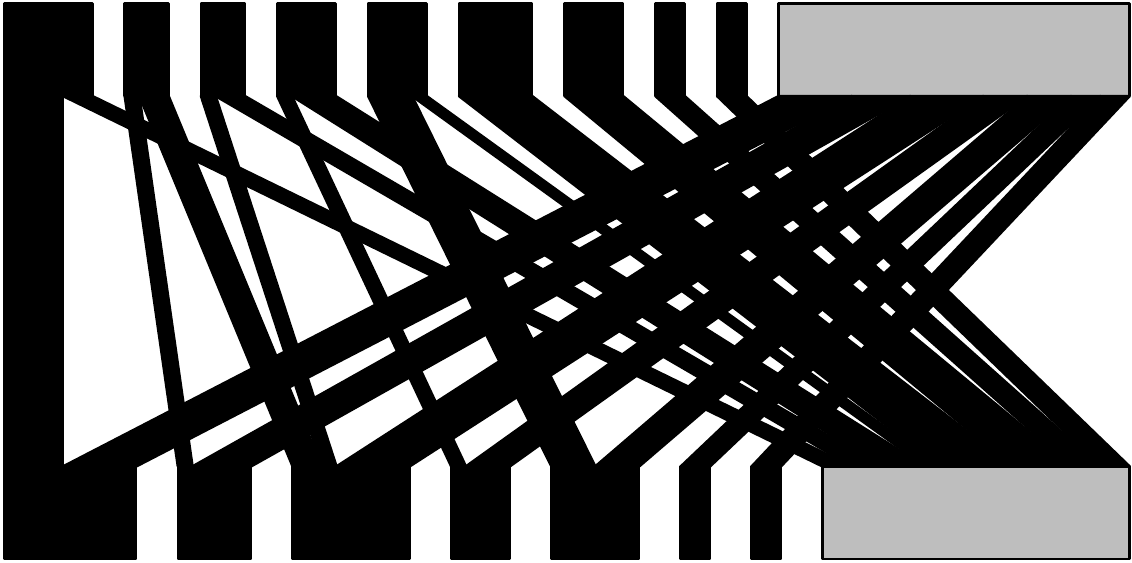}
        \caption{textile and leather \\ (N=123) ~ \\}
        \label{text}
    \end{subfigure}
        ~ \hfill
    \begin{subfigure}[b]{0.30\textwidth}
        \includegraphics[width=\textwidth]{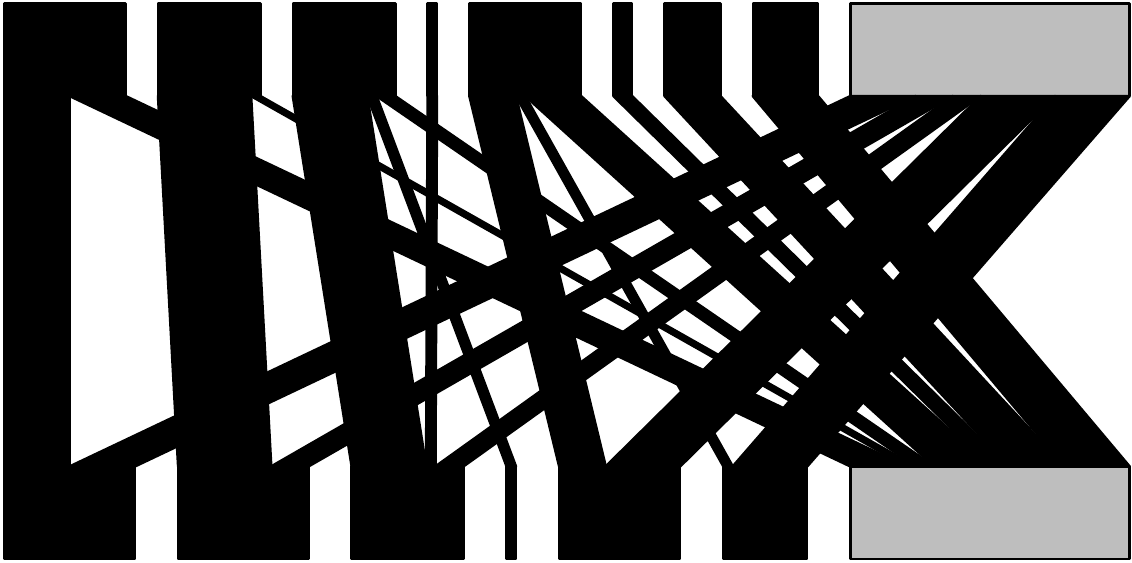}
        \caption{microbiology and immunology \\ (N=226)}
        \label{micro}
    \end{subfigure}
\end{figure}

\subsection{Explaining the stability of cores}

To analyse the differences in the stability of cores among scientific fields, Cugmas et al. \cite{cugmas2015} classified the fields into two categories: fields natural sciences and mathematics, engineering sciences and technologies, medical sciences, biotechnical sciences into the category "the natural and technical sciences" and social sciences and humanities into the category the "social sciences and humanitites". The selected features lowering the stability of cores were the splitting of clusters and out-of-cores researchers and therefore, the stability of clusters were measured by the MAWS2. They shows that there is no statistically significant difference in average core stability.

Given the high level of variability in the characteristics of the co-authorship networks and the blockmodel structures across scientific disciplines, the stability of the cores must be controlled by some additional network and blockmodel characteristics. Therefore, to explain the differences in the stability of cores across scientific fields, as controlling explanatory variables Cugmas et al. \cite{cugmas2015} also included in the linear model\footnote{The Least Squares Method was used to estimate the parameters' values. The correlations among independent variables were observed before the parameters were estimated. After that, the Variance Inflation Factor was checked to further investigate the potential problems of multicolinearity. The distribution of the residuals was also examined to identify the potential problems of heteroscedasticity or other un-satisfied assumptions.} the characteristics of the networks (number of researchers, growth from the first period to the second period in the number of researchers and the growth of the density) and the obtained blockmodels (average core size, percentage of cores, presence of a bridging core in the first time point, percentage of departures).

\begin{table}[ht!]
\caption{The impact of the characteristics of the network, blockmodel and disciplines on the stability of the cores}
\label{model1}
\centering
{\small
\begin{tabular}{lrrr|rrr}
  \hline
  & & Model 1 &                        &  & Model 2 &  \\  \hline
 & $b$ & $SE(b)$ & $p$                       & $b$ & $SE(b)$ & $p$ \\  \hline
  intercept                                                & 0.0906 & 0.2027 & 0.66 & 0.8349 & 0.1840 & 0.00 \\ 
   \hline
  \begin{tabular}[c]{@{}l@{}}number of researchers\\ (first time period) \end{tabular}                 &-0.0002 & 0.0003 & 0.58 & 0.0001 & 0.0002 & 0.77 \\ 
  \begin{tabular}[c]{@{}l@{}}growth of number of researchers\\ (1st and 2nd time period) \end{tabular} & 0.0010 & 0.0015 & 0.53 & 0.0004 & 0.0010 & 0.72\\ 
  \begin{tabular}[c]{@{}l@{}}growth of density\\ (1st and 2nd time period) \end{tabular}               & 0.0015 & 0.0010 & 0.04 & 0.0091 & 0.0005 & 0.07\\ 
  \begin{tabular}[c]{@{}l@{}}average core size\\ (1st time period) \end{tabular}                       & 0.0625 & 0.0177 & 0.00 & 0.0053 & 0.0152 & 0.73\\ 
  \begin{tabular}[c]{@{}l@{}}percentage of cores\\ (1st and 2nd time period) \end{tabular}             & -0.0054 & 0.0049 & 0.28 & -0.0069 & 0.0033 & 0.05\\ 
  \begin{tabular}[c]{@{}l@{}}presence of the bridge\\ (1st time period) \end{tabular}                  &  0.0404 & 0.0450 & 0.38 & -0.0005 & 0.0313 & 0.99\\ 
  percentage of out-of-cores                                                                             & \multicolumn{3}{c|}{\emph{not included}} & -1.0160 & 0.1667 & 0.00 \\
   \hline
  humanities (reference category) &  &  &  \\ 
   \hline
  natural science and math.                         &-0.1511 & 0.0892 & 0.10 & 0.0378 & 0.0680 & 0.58\\ 
  engineering sciences and tech.                          &-0.0120 & 0.0834 & 0.89 & 0.1339 & 0.0615 & 0.04\\ 
  medical sciences                                        &-0.0850 & 0.0954 & 0.38 & 0.1421 & 0.0748 & 0.07\\ 
  biotechnical sciences                                   &-0.0353 & 0.1008 & 0.72 & 0.0338 & 0.0694 & 0.63\\ 
  social sciences                                         &-0.0707 & 0.0844 & 0.41 & 0.0847 & 0.0626 & 0.19\\ 
   \hline
\hline
  \multicolumn{1}{l}{\emph{Number of obs. (disciplines):}} & \multicolumn{3}{r}{\emph{43}}                                   & \multicolumn{3}{r}{\hfill \emph{43}}   \\ 
  \multicolumn{1}{l}{\emph{Adjusted $R^2$:}}                       & \multicolumn{3}{r}{\emph{0.23}}                                 & \multicolumn{3}{r}{\hfill \emph{0.65}}   \\ 
  \multicolumn{1}{l}{\emph{F Statistics:}}                         & \multicolumn{3}{r}{\emph{2.151 (11; 31) ($p < 0.05$)}}     & \multicolumn{3}{r}{\hfill \emph{7.375 (12; 30) ($p < 0.01$)}}   \\ 
  \multicolumn{1}{l}{\emph{Method of estimation:}}                 & \multicolumn{3}{r}{\emph{Least Squares Method}}                 & \multicolumn{3}{r}{\hfill \emph{Least Squares Method}}   \\ 
   \hline
\end{tabular}
}
\end{table}

The main results are presented in Table~\ref{model1}. Here the humanities is used as the reference field since many studies suggest the social sciences are becoming more similar to the natural and technical sciences regarding publishing behavior \citep{kyvik2003, kronegger2015b}. In Table~\ref{model1} (Model 1), one can see there are no statistically significant differences between the humanities and other scientific fields when the percentage of departures is not included in the model. However, when the percentage of departures is included in the model, the differences in the mean stability of cores between the humanities and the engineering sciences and technologies and the humanities and the medical sciences become statistically significant ($p<0.10)$. Here, the scientific disciplines of both fields are seen as more stable than the humanities. Since the percentage of out-of-cores researchers forms part of the core stability index, the statistically significant differences between the mentioned scientific fields are mainly the consequence of the splitting of cores.

The effects of some controlling explanatory variables are statistically significant at ($p < 0.10$) as well. When the variable percentage of out-of-cores researchers is included in the model (see Model 1 in Table~\ref{model1}), the growth of the density and the average core size in the first time period is statistically significant. The density is defined as the share of all realized ties from all possible ties. The value is typically greater in the case of smaller networks with a low percentage of researchers in the periphery and many cores with a lot of researchers included. Therefore, together with the variable average core size, it can be argued that in the case of greater density there are more researchers who co-authored only occasionally (semi-periphery) and more complete cores with a higher number of researchers. The probability of creating ties with new researchers is therefore lower and the stability of the cores is higher. Similarly, De Haan et al. \cite{de1994a} mentioned that the size of a research group affects the persistence of collaboration.

When the percentage of out-of-cores researchers is included in the model, the growth of density and the percentage of core are statistically significant ($p<0.10$) along with the controlling explanatory variable percentage of out-of-cores researchers, which is highly statistically significant ($p < 0.01$). Since the latter is part of the definition of response variable, the percentage of explained variance of stability of cores is much higher in the model that includes percentage of out-of-cores researchers (Adjusted $R^2 = 0.65$) compared to the model where this variable is not included (Adjusted $R^2 = 0.23$).

\section{Conclusions}

It is crucial to understand how modern science works to ensure appropriate research and development policies are adopted that lead to improved scientific output. Modern information databases containing information about scientific bibliographic units can help in understanding the formation and maintenance of co-authorships among researchers. Although the borderline of scientific collaboration is unclear and there is no accurate way to measure it \citep{katz1997}, co-authorships can be seen as a rough operationalization of scientific collaboration, which is one of the primary results of scientific collaboration and represents one of the most formal manifestations of scientific communication \citep{groboljsek2014}.

The co-authorship patterns were studied through co-authorship networks. These are networks where the vertices present authors (or researchers) and a link between them exists if they co-authored at least one scientific bibliographic unit. Kronegger et al. \cite{kronegger2011} analyzed the co-authorship networks of four Slovenian scientific disciplines (physics, mathematics, biotechnology and sociology) in four periods (from 1990 to 2010). Only by observing the number of links among different scientific disciplines could they confirm that different co-authorship cultures exist between "lab" and "office" scientific disciplines. Publishing in co-authorship is more common in "lab" sciences while solo-authored scientific units are more common in "office" scientific disciplines where teamwork is not so crucial for the research. Hu et al. \cite{hu2014} classified four scientific disciplines in two groups: theoretical disciplines and experimental disciplines. They observed a stronger correlation between collaboration and productivity in experimental disciplines compared to theoretical ones.

However, one of the chief interests of the study by Kronegger et al. \cite{kronegger2011} was on the global network structure. To analyze this, they used generalized blockmodeling on network slices in four 5-year consecutive periods. They confirmed the network structure of multi-cores, semi-periphery, and periphery being present in all scientific disciplines. It can happen that the mentioned structure is not so outstanding at the earliest time points in some scientific disciplines. They defined the core as a group of researchers who very systematically co-author with each other, but who usually do not collaborate with researchers from the other cores. The semi-periphery consists of authors who collaborate with others inside the network, but in a less systematic way. It is not possible to cluster researchers from the semi-periphery into several well-separated clusters. The last part, the periphery, is the biggest part of the analyzed networks. These are authors who publish at least one bibliographic unit but as a single author or with researchers from abroad (with researchers not registered at the Slovenian Research Agency). Besides the main three types of mentioned positions, they observed so-called bridging cores. These are groups of researchers who collaborate with at least two other cores, which are not connected. 

Cugmas et al. \cite{cugmas2015} extended the analysis at the level of all Slovenian scientific disciplines. Like Kronegger et al. \cite{kronegger2011}, they analyzed data for the period between 1991 and 2010, but only analyzed the data in two 10-year long periods. The wider time span has an effect on the network density. Despite this, there are some scientific disciplines without any links in the first or second period, e.g. theology. These kinds of scientific disciplines were removed from the analysis, leaving 43 out of 72 scientific disciplines for further analysis. The assumed multi-core-semi-periphery-periphery structure was confirmed as being present in all analyzed scientific disciplines. In many of them, bridging cores are also found. On average, the number of researchers is increasing in time, also reflected in the higher average core size which is higher in the second period in both scientific disciplines from the fields of the natural and technical sciences and scientific disciplines from the social sciences and humanities. Here, the average size of cores is smaller in the social sciences and humanities in both time periods. The differences may be affected by the fact that authors from abroad are not included in the analysis since the rate of co-authored publications with researchers from abroad is higher in fields of the natural and technical sciences than in the social sciences and humanities. As reported by Kronegger et al. \cite{kronegger2011}, the main part of co-authorship networks is represented by authors from the periphery, which is generally decreasing over time. 

Another important property of co-authorship networks is that the cores can emerge in time, disappear, split, or merge. To measure the stability of cores, operationalized with these four rules in different ways, several indices were proposed. The value of each was calculated for each scientific discipline and, based on this, the scientific disciplines were clustered in three clusters. The observation of these clusters reveals that, according to the values of the proposed indices, they are mainly characterized by different levels of stability of the clusters and can therefore be ordered from least to most stable. The majority of scientific disciplines were classified in the stable-unstable cluster (22 scientific disciplines) while only a few were classified in the most stable cluster (8 scientific disciplines). It turns out that the average percentage of researchers classified in the cores in both periods is increasing along with the stability of the clusters. On the other hand, the percentage of researchers leaving the cores in the first time period and the percentage of researchers joining the cores in the second period is decreasing with the average stability of cores by the obtained clusters. The average core size is higher in the most stable cluster of scientific disciplines, indicating the existence of well-established scientific research teams in these scientific disciplines. De Haan et al. \cite{de1994a} mentioned that the size of a research group affects the persistence of collaboration.

A higher average number of researchers is associated with a lower level of stability of the cores. There are several explanations for this phenomenon, including the fact there are many opportunities to collaborate with different researchers in bigger scientific disciplines. The others are chiefly related to national research and development policies (e.g., the Young Researchers Program) and the nature of the work in such scientific disciplines (e.g., lab vs. office scientific disciplines or natural and technical sciences vs. social sciences and humanities).

To explain the differences between the natural and technical sciences and the social sciences and humanities, Cugmas et al. \cite{cugmas2015} performed a linear regression in which several network- and blockmodel-related variables (number of researchers in the scientific discipline, growth in number of researchers, growth in density, average core size, average percentage of cores, presence of a bridge) were included in the model as explanatory variables, while the stability of cores (response variable) was operationalized by the MAWS2, where the splitting of cores and out-of-cores researchers reduces the value of an index and thus indicates lower core stability. There were no statistically significant differences in the mean stability of cores between the natural and technical sciences on one hand and the social sciences and humanities on the other. This could be caused by many differences in the publication culture within these two groups of scientific disciplines (which is also a consequence of the characteristics of the particular national classification scheme of scientific fields, disciplines and sub-disciplines). In fact, even within some scientific disciplines the publication cultures vary widely. Moody \cite{moody2004} found that quantitative work is more likely to be co-authored than non-quantitative work in sociology.

However, when the analysis is performed on the level of scientific disciplines, the scientific discipline natural sciences and mathematics is statistically significantly ($p<0.10$) less stable than the field of the humanities. The growth of density and the average size of cores are also statistically significant ($p<0.05$) and positively correlated with the stability of the cores. When the additional variable percentage of out-of-cores researchers is included in the model, the difference in the average stability of cores between the humanities and medical sciences becomes statistically significant ($p<0.10$). Here, it must be highlighted that when the variable percentage out-of-cores researchers is included in the model, only the splitting of cores is seen as a features indicating lower core stability.


\bibliography{chapter}

\end{document}